\documentclass[pre, twocolumn,...]{revtex4-2}

\usepackage{amsmath}
\usepackage{graphicx}
\usepackage{dcolumn}
\usepackage{bm}

\begin{document}


\title{Medial packing and elastic asymmetry stabilize the double-gyroid in block copolymers}

\author{Abhiram Reddy}
\author{Michael S. Dimitriyev}
\author{Gregory M. Grason}
\affiliation{ 
Department of Polymer Science and Engineering, University of Massachusetts, Amherst, MA 01003
}%

\begin{abstract}
Triply-periodic networks are among the most complex and functionally valuable self-assembled morphologies, yet they form in nearly every class of biological and synthetic soft matter building blocks.  
In contrast to simpler assembly motifs -- spheres, cylinders, layers -- TPN assemblies require molecules to occupy variable local domain shapes, confounding attempts to understand their formation.
Here, we examine the double-gyroid (DG) network phase of block copolymer (BCP) melts, a prototypical soft self-assembly system, by using a geometric formulation of the strong stretching theory (SST) of BCP melts.
The theory establishes the direct link between molecular BCP packing, thermodynamics of melt assembly and the {\it medial map}, a generic geometric measure of the center of complex shapes.
We show that ``medial packing'' is essential for thermodynamic stability of DG in strongly-segregated melts, reconciling a long-standing contradiction between infinite- and finite-segregation theories, corroborating our SST predictions at finite-segregation via self-consistent field calculations.
Additionally, we find a previously unrecognized non-monotonic dependence of DG stability on the elastic asymmetry, the comparative entropic stiffness of matrix-forming to tubular-network forming blocks.
The composition window of stable DG -- intermediate to competitor lamellar and columnar phases -- widens both for large and small elastic asymmetry, seemingly overturning the heuristic view that packing frustration is localized to the tubular domains.  This study demonstrates utility of geometric optimization of {\it medial tesselations} for understanding in soft-molecular assembly. As such, the particular medial-based approach deployed here is readily generalizable to study packing frustration far beyond the case of DG morphologies in neat BCP assemblies.
\end{abstract}

\keywords{Suggested keywords}
\maketitle

Triply-periodic network (TPN) morphologies constitute ``natural forms'' of self-assembled soft matter.  Forming in nearly every class of amphiphilic molecular building blocks, from surfactants \cite{LUZZATI1967, Scriven1976, Sorenson2011} and lyotropic liquid crystals \cite{Chen2020} to complex shape amphiphiles \cite{Cho2004, Huang2015}, TPN domain structures most typically occur under conditions intermediate to those where cylindrical and layered morphologies, two other canonical forms of soft matter organization \cite{HYDE1997}, form.  By far the most commonly observed TPNs are the cubic domain networks, the double-gyroid (DG) double-diamond (DD) and double-primitive (DP).  Heuristically, the local structure of these ``in-between'' phases is often characterized by the shape of the intermaterial dividing surface (IMDS) that separates unlike components \cite{Thomas1988}, which is effectively more curved than planar layers, but less-so than cylinders.  Likewise, the global structure of TPNs reflects a hybridization between cylinders and layers \cite{Hyde2003}, with one domain forming a double network of interconnected tubes (e.g.~meeting at 3-, 4-, and 6-valent connections for DG, DD, and DP respectively), interspersed with a slab-like matrix layer, whose undulating shape approximates a triply-periodic minimal surface (e.g.~the Gyroid, Diamond, Primitive surfaces).
The combination of intercatenated, polycontinuous domain structures with complex (3D crystallographic) symmetries make TPN morphologies among the most sought after self-assembly architectures for functional nanostructured materials \cite{Lee2014, Hur2011} and their robust formation pathways make them natural photonic structures, endowing organisms from butterflies and beetles to birds with structural coloration \cite{Saranathan2010, Saranathan2021}. 

Despite their formation by diverse molecular systems, universal explanations for why various TPN structures form under equilibrium assembly remain elusive.
A common theme in most rationalizations of TPN formation is the concept of {\it packing frustration} \cite{Anderson1988, Matsen1996, GRASON2006, Schroder-Turk2007,ACShi2021}, which loosely refers to the incompatibility between a preferred local geometry for constituents and constraints of space filling.
Unlike cylinders or layers, which have uniform shape and thickness, IMDSs of tubular TPNs have variable curvature and molecules extending from this surface to the ``center'' of the domain have to reach variable distances, yet need to do so at nearly constant density.  
While this heuristic picture is widely held, a basic gap remains between specific geometric measures of packing and IMDS shape, molecular configurations of constituents, and the thermodynamic selection of TPN structures that form in equilibrium.

In this paper, we revisit the theory of self-assembly of a TPN morphology based on the so-called {\it strong-segregation theory} (SST) of AB block copolymer (BCP) melts.  BCP melt assembly is an ideal system for understanding complex morphology formation, owing to the fact that equilibrium behavior of variable chemical and architectural compositions are well captured by a self-consistent field (SCF) model \cite{Matsen2002}, dependent on only a few parameters: the composition fractions $f$ of component blocks; the $\chi$ parameter describing immiscibility between unlike components; and chain length $N$ ($\chi N$ controls the effective degree of segregation in assembled morphologies). Nonetheless, the understanding of the stable TPN phases of BCP melts has a complex and unsettled history \cite{Reddy2021}.  

\begin{figure*}[!htb]
\center
\includegraphics[width=0.85\textwidth]{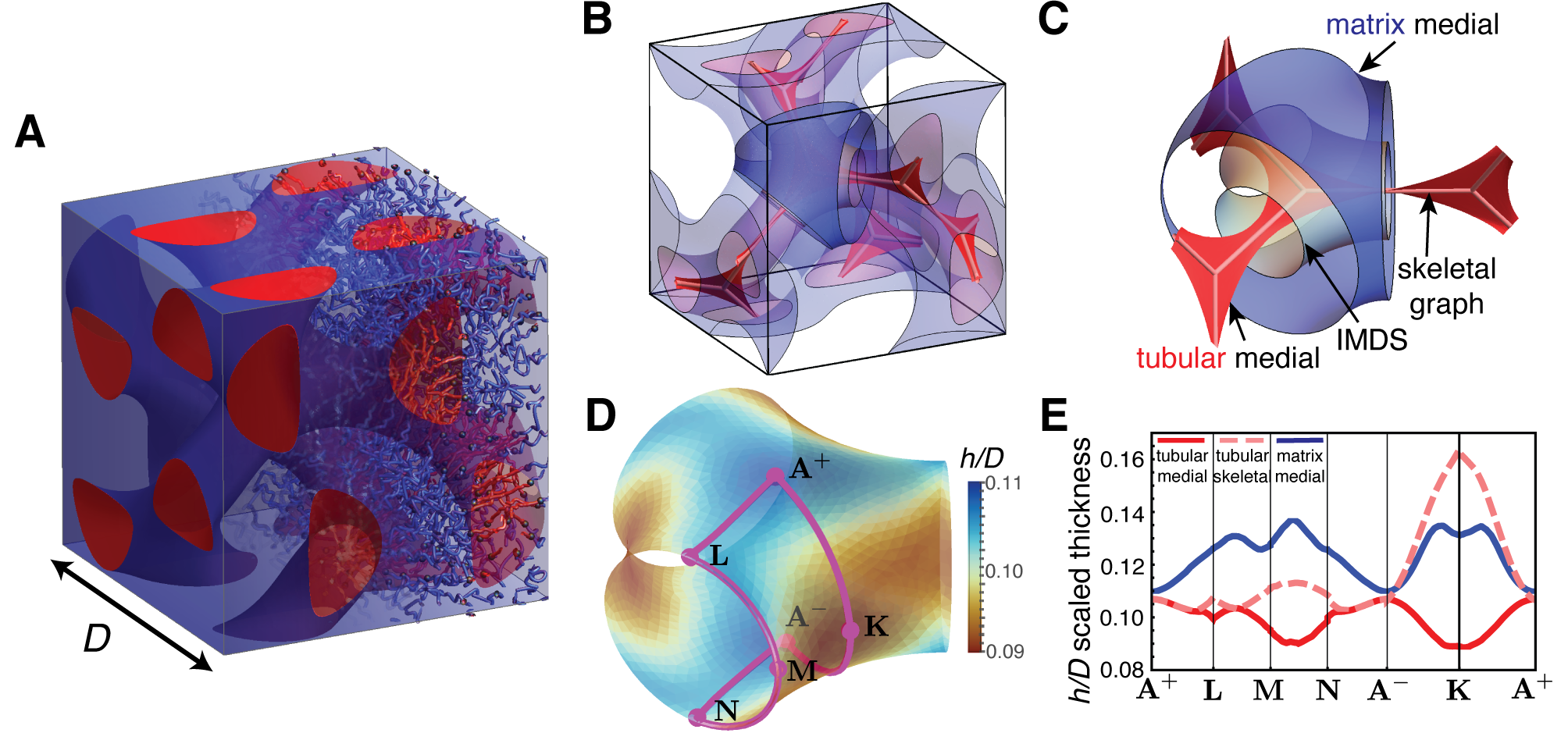}
\caption{\label{fig:1} Schematic of chain packing in tubular and matrix domains of self-assembled diblock copolymer double-gyroid (DG) morphology with cubic unit cell length $D$ is shown in (A), where red and blue regions show domains of A and B segments. (B) shows only a single-gyroid domain of the structure shown in (A) with red and blue surfaces corresponding to medial sets of single tubular and matrix domains and the skeletal graph shown as light red, with the highlighted nodal region shown in detail in (C). The spatial map of the tubular domain medial thickness is plotted on the nodal IMDS in (D), with a path bounding an asymmetric unit of the DG: 3- and 2-fold axes pass through points ${\bf A}^{\pm}$ and ${\bf K}$.  (E) shows a ``band diagram'' of local domain thickness $h$ as measured by distance to medial surface or skeletal graph.}
\end{figure*}

While initial experimental observations, based on AB starblocks, pointed to a DD ($Pn\bar{3}m$) network \cite{Thomas1986}, subsequent studies of linear AB diblocks \cite{Hajduk1994} identified the DG ($Ia\bar{3}d$) as the stable TPN (see Fig.~\ref{fig:1}A) at compositions intermediate to stable hexagonal cylinders (Hex) and lamellar (Lam) morphologies.  This was confirmed by computational SCF studies of triply-periodic morphologies \cite{Matsen1994}, showing that DG was the equilibrium TPN between Hex and Lam phases at low to moderate $\chi N$.  Extrapolations of phase boundaries to higher segregation first suggested that the stability window of DG would pinch off at sufficiently high $\chi N$ \cite{Matsen1996a}.  A parallel set of studies considered the stable TPN morphologies in the asymptotic limit $\chi N \to \infty$ based on SST \cite{Olmsted1994, Olmsted1998,Likhtman1994,Likhtman1997}, which computes the free energies directly from space-filling configurations of alternating, brush-like domains of BCP in competing arrangements.  Within SST, the thermodynamics become almost a purely geometric balance between entropic costs of chain extension and enthalpic cost proportional to the IMDS area \cite{GRASON2006}.  From this perspective, those stable phases in the SST limit might be viewed as ``natural forms'' of assembly, emerging from generic, geometric considerations.   Early SST calculations predicted that DG had the lowest free energy among competitor TPNs, but was nevertheless not stable relative to Hex and Lam as $\chi N \to \infty$~\cite{Olmsted1998, Likhtman1997}.  However, more recent experiments~\cite{Davidock2003} and SCF calculations~\cite{Cochran2006} that push to much higher $\chi N$ show that equilibrium DGs nevertheless persist to very high degrees of segregation.  Whether DG belongs to those natural forms stable in the $\chi N \to \infty$ limit, or otherwise relies on specific entropic corrections at finite segregation, has remained an open question~\cite{Lodge2020}. 

Here, we show that the resolution of this long-standing puzzle derives from a close connection between packing frustration in BCP and {\it medial geometry} of complex IMDS shapes.
Specifically, we generalize the SST approach to TPN structures and other complex BCP morphologies to incorporate degrees of freedom associated with the {\it terminal boundaries}~\cite{Reddy2021} between brush-like BCP domains and show that entropic considerations favor spreading of termini over the medial sets of TPN interfaces.  This medial packing not only leads to thermodynamic stability of the DG in the SST phase diagram, but further predicts a DG stability window that depends non-monotonically on the {\it elastic asymmetry} between blocks that compose the tubular and matrix domains.  
We further demonstrate the features of medial packing of subdomain chains that persist to finite degrees of segregation.
While we restrict our focus to the stability of one TPN morphology (DG) for a particular class soft molecular assembly (AB diblock melts), we posit that the basic paradigm and specific methodology for evaluating the thermodynamic costs of packing frustration extend to other complex morphologies assembled from a much broader class of soft matter building blocks.  

\section{Medial anatomy of TPN morphologies} 
To understand the thermodynamic costs of packing in TPN morphologies, it is necessary to measure how far chains must extend from the IMDS into the ``center'' of complex domains.  In BCP melts, these centers, defined as {\it terminal boundaries}~\cite{Reddy2021}, are locally 2D sets of contact points between opposing brush-like domains from distinct regions of the the IMDS. The length $h$ of the molecular extension from the IMDS to the terminal boundary defines a local {\it thickness}. 

Historically, the focus on packing frustration in TPN morphologies of neat (solvent-free) amphiphiles like BCP, as well as ``Type I'' lyotropic TPNs~\cite{Schroder-Turk2006}, has been on the tubular, double-network domains (e.g.~the red A domain in Fig.~\ref{fig:1}A).  Prior theories of TPN thermodynamics in BCP melts associated the centers of tubular domains with their {\it skeletal graphs}~\cite{Olmsted1998, Likhtman1997, Prasad2018}, 1D graphs that thread between adjacent $Z$-valent interconnections.  Figure 1B shows the skeletal graph within one of the two single-gyroid domains in the DG morphology, which connects the node centers located at 8 of the 16b Wyckoff positions of $Ia\bar{3}d$.  Focusing on a single 3-valent nodal region in Figure 1C, the largest distance from the IMDS (grey) to the skeletal graph (the maximal {\it skeletal thickness}) is associated with points from the saddle-like regions between struts that have to extend towards the center of the node.

Following an argument proposed by Schr\"oder-Turk, Fogden and Hyde for lyotropic morphologies and block copolymers \cite{Schroder2003, Schroder-Turk2006, Schroder-Turk2013}, it was shown~\cite{Reddy2021} that a more natural definition of local thickness is provided by the {\it medial map}, the set of points closest to a given point on a generating surface (or set of surfaces).  Points on the generating surface are mapped to {\it medial sets} by following the local surface normal until it intersects with another, remote surface normal at an equidistant point; the distance to those points $h$ is the {\it medial thickness}~\cite{Schroder2003,Reddy2021}.
This construction rigorously minimizes the distance map of {\it all points} enclosed within the volume bounded by the generating surface.  Thus, the medial map can be viewed as a generalization of \textit{centroidal} Voronoi cells \cite{Du1999}, in which distances to a set of generating points are minimized, to case of {\it generating surfaces}. Figs.~\ref{fig:1}B,C also show the medial surfaces for both the outer (matrix) and inner (tubular) domains for a level set model for DG.  
The outer medial surface, which divides the DG into two enantiomeric single gyroid domains, closely resembles the Gyroid (G) minimal surface, while the inner medial set is composed of a web-like surface that locally spreads out in the plane of the 3-fold junctions, and twists by $70.5^\circ$ from one node to the next.

Figure \ref{fig:1}D plots a spatial map of the medial thickness of the (inner) tubular domain on the nodal region of the IMDS of the DG model.
In the Fig.~\ref{fig:1}E ``band diagram,'' we compare domain thickness measures along a closed path passing through symmetry points of the nodal region. Notably, the magnitude and variation of skeletal thickness is not only greater than medial thickness, but the spatial pattern of thickness variation is nearly opposite between these measures.  
While the quasi-planar points $\mathbf{A^{\pm}}$ correspond to minimal skeletal thickness, these are points of maximal medial thickness.  
This contrast is even greater at the saddle-like ``elbows'' at 2-fold points {\bf K}, where the maximal skeletal thickness is nearly double the medial thickness, which is at a local minimum.
Contrary to the standard heuristic that the cost of packing frustration in TPN phases is confined to tubular domains, the medial thickness in the matrix domain (also shown in Fig.~\ref{fig:1}E) exhibits variability comparable to the tubular domain. These geometric observations strongly imply that theories based on the assumption of skeletal packing (i.e.~terminal boundaries lie along the 1D skeleton) dramatically overestimate the costs of chain stretching in the tubular block relative to medial packing (i.e.~ terminal boundaries lie along the medial set). As medial geometry encodes the shortest distance to the ``center'' of a complex morphology, previous studies have proposed that medial thickness can be used as a heuristic measure of packing frustration~\cite{Schroder-Turk2007, Schroder-Turk2013, Reddy2021} or otherwise an ingredient in phenomenological models of its costs~\cite{Chen2017} in TPN morphologies. 
We next exploit a fully molecular description of TPN assembly, the SST of BCP melts, to test and establish the direct connections between the medial geometry of complex TPN, the underlying configurations of molecular constituents and the thermodynamic stability of the DG phase.

\begin{figure}
\center
\includegraphics[width=0.425\textwidth]{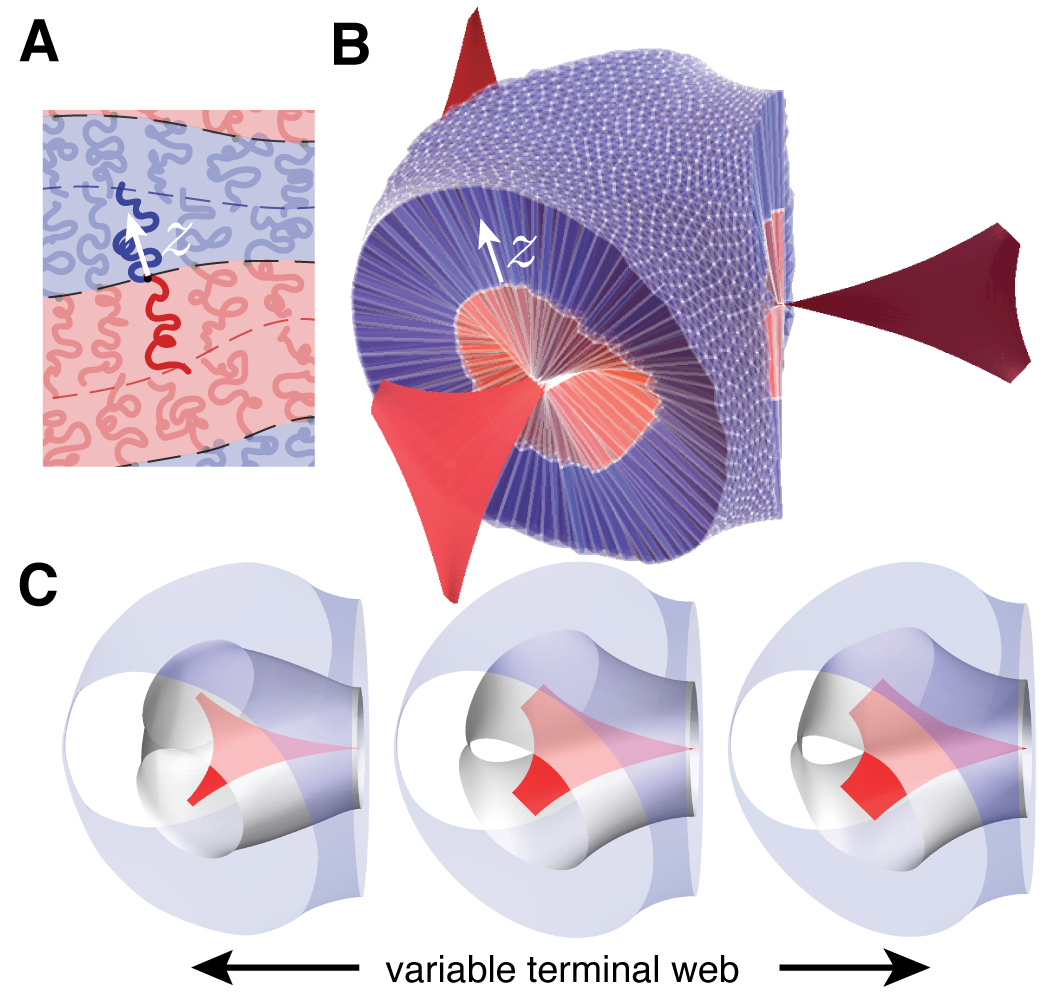}
\caption{\label{fig:2} Schematic of BCP packing within a variably-shaped morphology is shown in (A) where $z$ labels local extension of blocks from the IMDS.  (B) shows a (volume balanced) tessellation of the nodal region of DG, generated via the medial SST (mSST) construction.  (C) shows a sequence of mSST geometries (at fixed composition $f = 0.3$) highlighting lateral spreading of terminal webs achieved via parametric variation of DG generating surfaces (as described in the text). }
\end{figure}

\section{Thermodynamics of medial packing in double-gyroid}
To assess the importance of terminal spreading along medial surfaces, we turn to the SST of DG (and its competitor phases Hex and Lam)\cite{Semenov1985, Olmsted1998,Likhtman1994}.
Each BCP consists of $N$ segments, a fraction $f$ of which are A-type, with segment lengths $a_{\rm A}$ and $a_{\rm B}$.
As shown schematically in Fig.~\ref{fig:2}, microphase segregation partitions space into A and B domain, delineated by an IMDS, from which extended blocks form brushes up to a thickness defined by the terminal boundary.
Critically, as a solvent-free melt, volumes occupied by chains must obey a {\it local volume balance}: the ratio occupied by A:B blocks extending from each point on the IMDS is $f/(1-f)$.  
Given a volume-balanced space partition for morphology $X$, the free energy per chain (in units of $k_{\mathrm{B}}T$) is 
\begin{equation}
    F(X) = \frac{N \rho^{-1} }{V(X)}\Big [ \gamma A(X) + \frac{\kappa _{\rm A}}{2} I_{\rm A}(X) + \frac{\kappa _{\rm B}}{2}I_{\rm B}(X) \Big] 
    \label{eq: F}
\end{equation}
where $\rho V(X)/N$ is the number of chains in the assembly with total volume $V(X)$ and volume per segment $\rho^{-1}$ (see {\it SI Appendix, S1}). 
Remarkably, SST samples the relevant conformational fluctuations of underlying BCP molecules in the $\chi N \to \infty$ limit~\cite{Goveas1997}, and yet the free energy reduces to purely geometric measures of competing morphologies~\cite{GRASON2006}. 
The first term in eq.~(\ref{eq: F}) encodes the surface energy of AB contact, where $A(X)$ is the total area of the IMDS and $\gamma\propto \sqrt{\chi}$ is effective surface tension. 
The second and third terms encode the respective entropic costs of stretching Gaussian chains~\cite{Likhtman1994} within the \textit{parabolic brush} theory (PBT),
\begin{equation}
    I_{\alpha} = \int_{V_\alpha}{\mkern-10mu} \mathrm{d}V\, z^{2} \, , \, \text{with } \alpha = {\rm A}, {\rm B}
    \label{eq: I}
\end{equation} 
where $z$ is the chain extension from the IMDS (see Fig.~\ref{fig:2}2A).
Coefficients $\kappa_{\alpha}$ ({\it SI Appendix, eq.~(S4)}) describe molecular and architectural features of the blocks that control the effective entropic stiffness of the domains\footnote{While the PBT assumes free ends to be distributed throughout brush, this is violated for convex brush curvatures~\cite{Ball1991}; we show ({\it SI Appendix, S10}) that corrections to PBT for the variable convex domain shapes involved~\cite{Dimitriyev2021Arxiv} have negligible impact on the stability of TPNs.}.  


Per eq.~(\ref{eq: I}), maximizing entropy of component blocks is equivalent to minimizing second-moments of volume measured relative to the IMDS.
BCP packing is, in this sense, a variant of the quantizer problem~\cite{Du1999}, which seeks tessellations that minimize the second-moment of distance from a set of generating points, generalized to measure distance with respect to generating surfaces.
Given an IMDS shape, the medial map would provide the optimal $I_\alpha$ for each domain considered in isolation.
However, as shown in {\it SI Appendix, S3}, the medial map generally fails to satisfy the local volume balance constraint.
Moreover, stable morphologies also optimize the competition between IMDS area minimization and block stretching. 

To model this thermodynamic competition and test how closely physical assemblies come to realizing medial packing, we apply the following \textit{medial SST} approach to DG morphologies (see {\it SI Appendix, S3-S4}).
We begin with a broad set of generating surfaces ${\bf G}$ with $Ia\bar{3}d$ symmetry; here we have chosen ${\bf G}$ as level sets of a three Fourier-mode expansion of the gyroid symmetry.
From the medial map of ${\bf G}$ (for the double network) we generate the terminal boundaries of both domains, ${\bf T}_{\rm A}$ and ${\bf T}_{\rm B}$, referring to the tubular and matrix domains respectively.
The induced map between ${\bf T}_{\rm A}$ and ${\bf T}_{\rm B}$ approximates the mean trajectories of domain-spanning chains, whose shapes are locally approximated by triangular prismatic wedges, thus providing a complete tessellation of the DG structure.
The location of the IMDS is then determined locally along each trajectory according to the constraints of volume balance and chain composition, resulting in mutually compatible terminal boundaries, IMDS, and trajectories, as shown in Fig.~\ref{fig:2}A, from which we compute the geometric contribution to the free energy in eq.~(\ref{eq: F}).  
As shown in Fig.~\ref{fig:2}C, variation of the generating surface leads to variation in compatible IMDS shapes and variable spreading of the terminal boundaries for fixed block composition $f$, which we then optimize over in our medial SST approach (see {\it SI Appendix, S4}).
Notably, local volume balance generally requires some measure of tilt between trajectories and the IMDS, a signature of the deviation from strictly medial packing which we return to below.

\begin{figure}
\center
\includegraphics[width=0.4\textwidth]{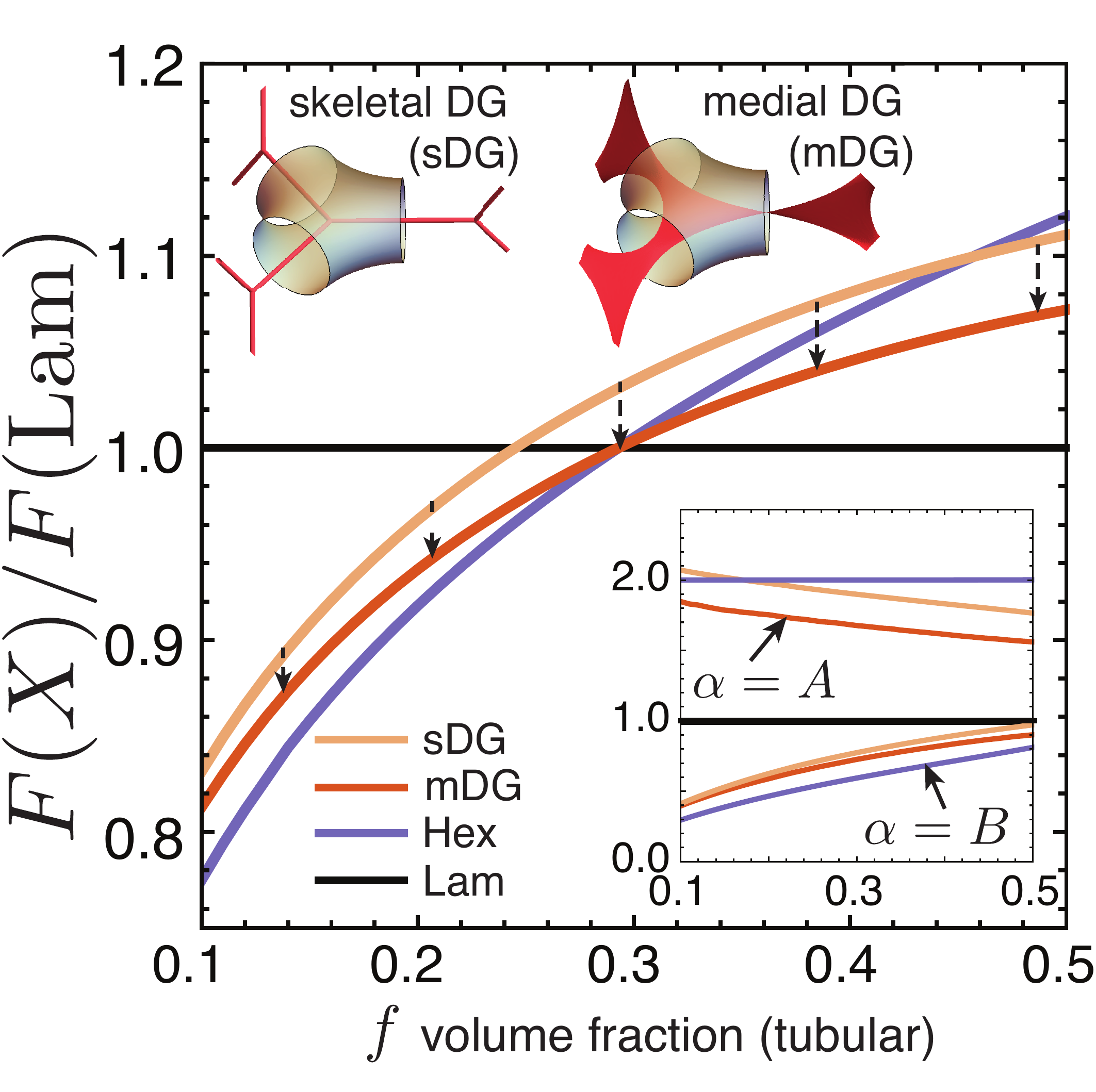}
\caption{\label{fig:3} Comparison of medial DG (mDG) and skeletal DG (sDG) to competitor hexagonal cylinder (Hex) and lamellar (Lam) phases, for conformationally symmetric linear diblocks.  Inset shows scaled entropic cost of stretching for both blocks (compared at equal IMDS area per chain).}
\end{figure}

\begin{figure*}
\center
\includegraphics[width=0.8\textwidth]{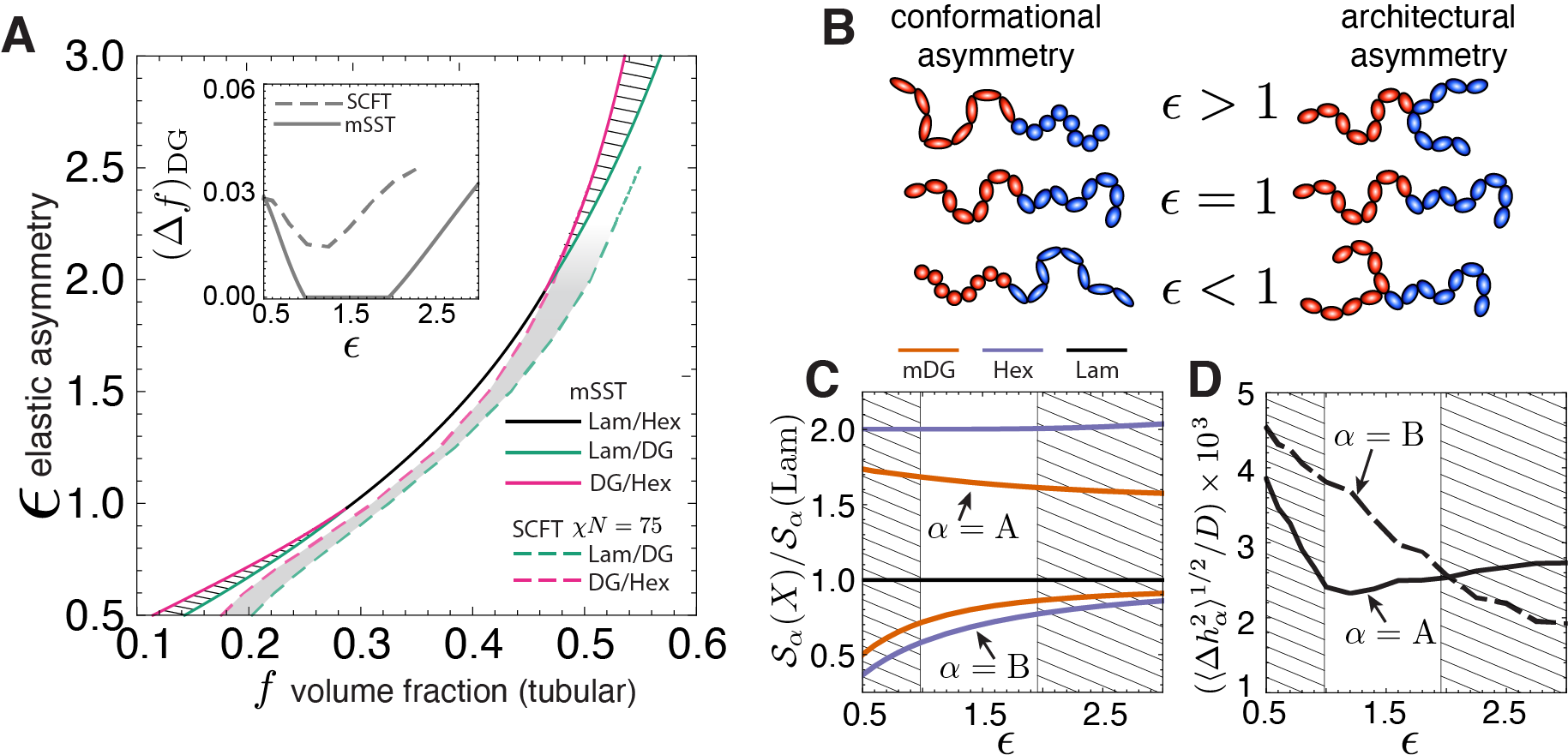}
\caption{\label{fig:4} (A) shows the stability of DG, Hex and Lam in the $f$-$\epsilon$ plane for mSST, with hatched regions indicating stable DG windows.  Dashed boundaries (and shaded grey) indicate corresponding boundaries for the SCF at $\chi N = 75$, and the inset compares the composition width of the DG stability region $(\Delta f)_{\rm DG}$ between infinite- and finite-segregation theories. (B) schematic illustration of the dependence of \textit{elastic asymmetry} on conformational (segment length) asymmetry and architectural (branch) asymmetry for AB miktoarm copolymers.   (C) compares the block stretching costs between the competitor morphologies along the line $F ({\rm Lam}) = F ({\rm Hex})$ for variable elastic asymmetry, while (D) compares the variance in domain thickness in DG domains along the same line.  In both, the hatched regions shown predicted DG windows.}
\end{figure*}

In Fig.~\ref{fig:3} we compare the free energy of DG to its competitor phases (Lam and Hex) for conformationally symmetric (i.e.~chains of equal segment lengths $a_{\rm A}$ and $a_{\rm B}$) linear AB diblocks, using a variable-IMDS shape SST for Hex~\cite{Grason2004}.   We also compare the medial SST construction of DG (mDG) to a skeletal construction (sDG) that is based on and closely approximates Milner and Olmsted's skeletal SST~\cite{Olmsted1998} for TPN phases (see {\it SI Appendix, S5}), showing the free energy per chain in medial packing is significantly lower ($\approx 2-4\%$) than skeletal packing.
In the inset Figure~\ref{fig:3}, we plot ${\cal S}_\alpha (X) = I_\alpha (X) A^2 (X)/V^3(X)$, a dimensionless measure of entropic stretching which compares block entropies at an {\it equal} IMDS area per chain (i.e.~fixed interfacial cost, see {\it SI Appendix, S2}).  
This shows that the free energy drop of mDG relative to sDG primarily derives from the reduced stretching in the tubular block, consistent with the reduced stretching of A blocks near point {\bf K} defined in Fig.~\ref{fig:1}D,E.
The entropic gain due to medial packing brings  DG into near-degeneracy (to within free energy 0.02\% excess) with its competitors at the boundary between Lam and Hex at $f \simeq 0.29$.  Next, we consider how elastic asymmetry between the domains lifts that degeneracy.

\section{Elastic asymmetry and double-gyroid stability}
Beyond the volume fraction $f$ of tubular blocks, stability of BCP phases is known to be critically dependent on chain architecture and conformational rigidity.
The \textit{elastic asymmetry} parameter $\epsilon \equiv (n_{\rm B} a_{\rm A})/(n_{\rm A} a_{\rm B})$, as defined by Milner for A$_{n_{\rm A}}$B$_{n_{\rm B}}$miktoarm stars~\cite{Milner1994} and illustrated in Fig.~\ref{fig:4}A, encodes the relative stiffness between blocks of unequal segment length $a_{\rm A}\neq a_{\rm B}$ (conformational asymmetry) and number of branches $n_{\rm A} \neq n_{\rm B}$ (architectural asymmetry).
In terms of eq.~(\ref{eq: F}), $\kappa_{\rm B}/\kappa_{\rm A} \propto \epsilon$, so that when $\epsilon > 1$, the B matrix domains are stiffer compared to the A tubular domains~\cite{Grason2004, GRASON2006}.
Notably, elastic asymmetry in sphere-forming BCP has been understood to unlock a host of ``self-alloying'' Frank-Kasper crystal structures through the amplification of packing frustration in the coronal blocks \cite{GRASON2006,Xie2014,Reddy2018,Dorfman2021}.

Figure~\ref{fig:4}B shows the phase boundaries between Lam, Hex and DG based on the mSST, in the $f$-$\epsilon$ plane. 
Surprisingly, the thermodynamic stability of DG exhibits a non-monotonic dependence on elastic asymmetry.
For intermediate values of asymmetry $0.98 <  \epsilon < 1.96$ there is no stable DG phase, and only a line of direct order-order transitions between Hex and Lam.  
However, for sufficiently high or low values of asymmetry (i.e.~$\epsilon \geq 1.95$ and $\epsilon \leq 0.98$) stable windows of DG open up intermediate to Hex and Lam\footnote{SST based on skeletal ansatz~\cite{Olmsted1998} showed previously that DG becomes stable for the especially large elastic asymmetry $\epsilon \gtrsim 9.5$.}.

For comparison, we show in Fig.~\ref{fig:4}B phase boundaries at finite segregation ($\chi N = 75$) from SCF calculations~\cite{Arora2016}.
Aside from a slight offset to larger $f$ (attributed to finite-$\chi N$ effects), the skewing of phase boundaries to larger $f$ with increasing $\epsilon$ agrees with mSST predictions. 
Notably, the width $(\Delta f)_{\rm DG}$ of stable DG compositions exhibits a non-monotonic dependence on $\epsilon$, widening for large and small asymmetry values, as shown in the inset of Fig.~\ref{fig:4}B.  
This suggests that the coupling between elastic asymmetry and packing frustration captured by mSST persists even at finite $\chi N$.

\begin{figure*}
\center
\includegraphics[width=0.9\textwidth]{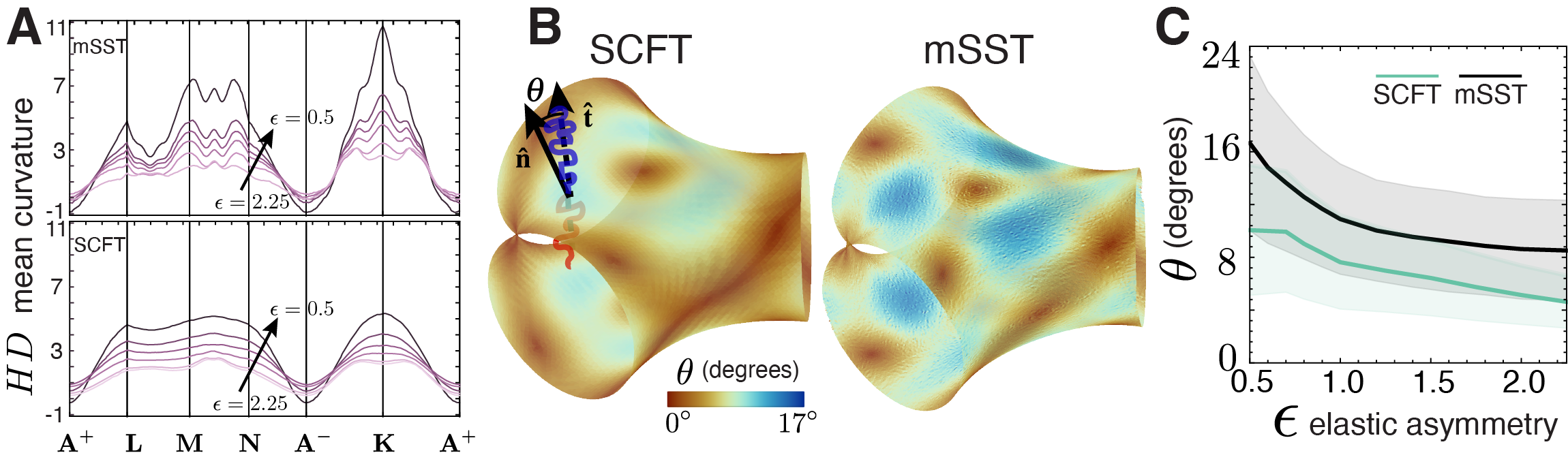}
\caption{\label{fig:5}(A) Band diagrams showing local mean curvature \textit{H} of DG IMDS in mSST and SCFT along the path defined in Fig.~\ref{fig:1}D. Maps of chain tilt at the IMDS for SCFT and mSST at $\epsilon = 1$ and $f= 0.3$ are shown in (B) and the corresponding dependence of tilt-angle distributions for DG morphologies along the $F({\rm Lam})=F({\rm Hex})$ line.}
\end{figure*}

We consider the mechanism underlying the non-trivial dependence of DG on elastic asymmetry, based on the dimensionless entropic costs ${\cal S}_\alpha (X)$ of competing phases plotted along the line $F({\rm Lam})=F({\rm Hex})$ for variable $\epsilon$ in Fig.~\ref{fig:4}C.
The stability of DG for $\epsilon > 1.96$ is consistent with the long-held notion that packing in the tubular block is problematic~\cite{Matsen1996}, so that discounting its thermodynamic cost relative to the matrix should increase its stability.
There is a more subtle scenario in the Hex phase, where the coronal domain is increasingly frustrated with increasing $\epsilon$, causing the IMDS to warp hexagonally~\cite{Grason2004}, which in turn deforms the core domain, as shown in the relative increase of entropic costs of the A block with $\epsilon$ (see {\it SI Appendix, S6}).
In contrast, terminal spreading in the tubular block of DG leads to a relative decrease of those entropic costs with $\epsilon$, allowing this phase to overtake Hex for sufficiently stiff matrices.

The ``reentrant'' stability of DG for $\epsilon \lesssim 0.98$ is more confounding, as the increased cost of packing frustration in the tubular block would seemingly destabilize the DG, according to the prevailing heuristic that suggests packing frustration is localized to the tubular block~\cite{Matsen1996}.
Instead we observe that the favorable A block stretching of DG relative to Hex is maintained and the nominal thickness of the tubular domain shrinks with composition along the $F({\rm Lam})=F({\rm Hex})$.  
This suggests the optimal DG morphology is able to redistribute the cost of packing frustration from the tubular domain to the matrix domain, in which thermodynamic costs are relatively less costly for $\epsilon <1$.
This is consistent with comparisons of the r.m.s.~block thickness in the A and B domains, plotted in Fig.~\ref{fig:4}D.
While DG exhibits a lower thickness variation in the matrix blocks for $\epsilon \gtrsim 2$, as $\epsilon$ is decreased, the optimal DG morphology reorganizes to reduce the thickness variation in stiffer, tubular domain.
Taken together, mSST predicts a nuanced picture of packing frustration in DG, in which terminal spreading along medial surfaces facilities a remarkable degree of adaptability to a broad range of molecular structure.

\section{Fingerprints of medial packing at finite segregation}
Medial SST predicts that optimal arrangements of DG assemblies rely on a variable spreading of BCP chain termini along medial surfaces interior to the domains. Here, we consider morphological features of DG assemblies and their variation with elastic asymmetry that derive from medial packing in the $\chi N \to \infty$ limit, but persist at finite segregation.  In Figure \ref{fig:5} we compare DG morphologies modeled by SCF at $\chi N = 75$ along the points ($\epsilon, f$) where $F({\rm Lam})=F({\rm Hex})$ according to mSST calculations.  Notably, SCF calculations include chain fluctuation effects associated with finite IMDS widths and contact zones between opposing brushes, without {\it a priori} assumptions about the underlying chain configurations.  In Figure \ref{fig:5}A, we compare ``band diagrams'' of the spatial map of IMDS mean curvature $H$ (along the path introduced in Fig.~\ref{fig:1}D) for a range of elastic asymmetry, $0.5<\epsilon<2.25$. These show qualitatively similar spatial patterns between mSST and SCF, with characteristically flatter regions along the 3-fold directions (${\bf A}^{\pm}$) and maximal curvature at the elbows (${\bf K}$), although magnitudes and fine features differ between infinite- and finite-segregation models (analogous comparisons are shown for Gaussian curvature in {\it SI Appendix, Fig.~S9A,B}).  More significantly, the variance of mean curvature is shown to decreases systematically with increasing elastic asymmetry (see {\it SI Appendix, Fig.~S9C}), suggesting that IMDS shapes tend more towards area-minimizing \cite{Anderson1988, Reddy2021} (i.e.~constant mean curvature) shapes as the ratio of matrix to tubular block stiffness grows.

Underlying the adaptation of the DG geometry is the increasing lateral spread of the terminal webs of the tubular blocks with increasing elastic asymmetry, shown in {\it SI Appendix, Fig.~S6}.  This trend might suggest a discernible measure of displacement of free A block ends from the skeletal graph for increasing $\epsilon$.  In {\it SI Appendix, Fig.~S13}, we compare the distance distributions of free A block ends from the skeletal graphs for three values of $\epsilon$.  While there is indeed a systematic shift of end distributions away from the skeleton observed for increasing $\epsilon$ in both SCF and SST results, free ends are always well-distributed throughout brushes that any signal of medial spreading in chain ends is rather diffuse.

For an alternative molecular fingerprint of the underlying terminal boundaries, we turn to chain orientation.  As noted above, in the mSST construction, local volume balance extending away from the same points on the IMDS requires some measure of chain tilting from the local IMDS normal, a signature of deviation from strictly medial packing.  Using the polar order parameter extracted from SCF~\cite{Prasad2017} and the direct trajectories from optimal mSST configurations, we compare the map of the local tilt angle $\theta$ relative to the IMDS at finite-$\chi N$ and $\chi N \to \infty$ in Fig.~\ref{fig:5}B.  Both show strikingly similar spatial patterns of tilted and normal regions.  Although finite-$\chi N$ results show a smaller magnitude of tilt, both SCF and mSST show near-identical patterns of normal packing ($\theta = 0$).  This occurs for the quasi-planar 3-fold points (${\bf A}^{\pm}$) along the elbow (extending from ${\bf N}$ to ${\bf K}$) and also at a satellite spot along the strut (between ${\bf L}$ and ${\bf M}$).  More careful analysis in {\it SI Appendix, Fig.~S12} shows that this satellite spot is located precisely along the ray that is co-normal to the underlying terminal web, and hence the rotation of that normal zone reflects (half of) the $70.5^\circ$ twist between nodes in DG.  In this way, we observe the non-trivial pattern of tilt at the IMDS as an indirect image of the terminal packing within the domain.  

Figure~\ref{fig:5}C shows the magnitude and variation of the tilt distribution as function of elastic asymmetry.  While magnitudes of tilt are larger for infinite- compared to finite-segregation theories, both show a consistent trend decreasing from relatively high tilt for $\epsilon < 1$ to approach a more normal (i.e. $\theta \to 0$) packing for $\epsilon >1$.  This latter tendency towards normal orientation implies that chain packing approaches closer to the medial limit as matrix domains stiffen and stable DG morphologies shift to larger A-block compositions, ultimately, becoming majority tubular structures for $\epsilon \gtrsim 2.25$. 

\section{Discussion}

Our results establish the relevance of medial geometry to the molecular degrees of freedom of TPN assembly of BCP melts. While constraints of local volume balance in this neat system frustrate perfect medial packing (i.e.~normal to IMDS) of chains, variable spreading of terminal ends of blocks along the medial surfaces nonetheless is essential to the stability of DG in the $\chi N \to \infty$ limit.  Moreover, we observe that the degree conformity with perfect medial packing for DG itself varies with molecular parameters controlling the ratio of domain stiffnesses.  In this light, the mSST reformulates the notion of packing frustration for BCP assembly, and we anticipate implications well beyond the DG stability in melt assembly.

First, the mSST approach extends straightforwardly to other TPN morphologies, such as the cubic cousins of DG -- DD and DP.  Indeed, it can be shown that mSST variants of DD and DP exhibit lower free energies than the prior skeletal ansatz. However, over the range of conditions presented here for AB BCP melts, the free energies of DD and DP always exceed that of DG.  As we will describe elsewhere, we expect that the  stability of DG over these competitors derives largely from uniquely smooth geometry of terminal webs of DG.

Beyond TPN phases, it can be expected that medial packing motifs play a key role in other complex BCP morphologies, notably the Frank-Kasper (FK) phases of sphere-like domains that are stabilized at high $\epsilon$~\cite{Dorfman2021}.  It is well appreciated that phases ``self-alloy'' into multiple symmetry- and volume-inequivalent domains.  Crudely speaking, the terminal boundaries between coronal blocks are polyhedral cells, similar to Voronoi cells.  However, medial analysis of FK-forming domains suggests that the terminal packing in core domains is more subtle \cite{Reddy2021}.  These FK structures (e.g.~A15 and C15) are composed of populations of more isotropic domains, with inner medial sets localized around points, mixed with another population of more anisotropic domains possessing planar, disk-like medial sets.  Prior SST models for explaining FK phase formation are based on the evidently oversimplified assumption that core termini are localized to the centroids of domains~\cite{GRASON2006,Reddy2018}, and hence likely considerably overestimate the entropic costs in these anisotropic domains, whose local packings include quasi-lamellar regions.  We anticipate that properly accounting for terminal spreading via an extension of mSST may be critical for understanding open questions in FK formation. For example, what controls the sequence of stable sphere phases (BCC$\to \sigma \to$ A15) found when increasing elastic asymmetry and core block fraction \cite{Xie2014,Dorfman2021}? 

Lastly, we briefly comment on likely effects on medial packing beyond neat (i.e.~melt) systems.  In blends, adding low-molecular weight species that dissolve in either one or both BCP blocks (i.e.~the so-called wet brush regime) will certainly have the effect of introducing {\it compressibilty} to those solvated domains.  In the context of medial packing specifically, compressibility has the effect of relaxing the local volume balance constraint, and thereby should reduce the frustration of medial packing.  Hence, we speculate that a direct consequence of blending with low-molecular weight solvent species should be a progression toward untilted (i.e.~medial) packing, which may establish a convenient conceptual link between medial packing frustration in neat (thermotropic) and concentrated (lyotropic) phases.  Mixtures of BCP with high-molecular weight species (e.g.~homopolymer of A or B type) are generally expected to lead to local segregation of additives from the molten brushes of the BCP~\cite{Matsen1995a, Matsen1996}, in effect sequestering them into interstitial ``hot spots'' between those brushes, in turn relaxing the free energy cost of filling space for those local regions.  SCF results~\cite{Matsen1995a,Martinez-Veracoechea2009}, and experiments~\cite{Takagi2017}, show that the stable TPN morphology can be altered upon blending homopolymer to the tubular domain, from DG to DD, and potentially to DP upon further homopolymer fraction.  Analogous effects altering the symmetry of the stable morphology have been observed and studied theoretically in FK-forming sphere phases of BCP and homopolymer blends~\cite{Mueller2020, Cheong2020}.  While it is arguably intuitive that homopolymer blends relax packing frustration in complex phases, it remains to be understood what geometric features of these morphologies lead a complex ``host'' to gain or lose stability over another upon loading with a ``guest'' species.  We anticipate that hot spots can themselves be directly related to the medial geometry of a complex morphology, and moreover, the mSST construction may be readily extended to provide a direct map of the location and thermodynamics of these hot spots upon blending.

\begin{acknowledgments}
The authors are grateful to E. Thomas, G. Schr\"oder-Turk, F. Bates, K. Dorfman and S. Park for stimulating discussions and valuable comments  on this work.  This research was supported by the US Department of Energy (DOE), Office of Basic Energy Sciences, Division of Materials Sciences and Engineering under awards DE-SC0014599 and DE-SC0022229. Medial SST (mSST) and SCFT calculations were
performed on the UMass Cluster at the Massachusetts Green High Performance Computing
Center
\end{acknowledgments}

\bibliography{RevTex_MedialSST}

\end{document}



\title{Supporting Information: Medial packing and elastic asymmetry stabilize the double-gyroid in block copolymers}

\author{Abhiram Reddy}
\author{Michael S. Dimitriyev}
\author{Gregory M. Grason}
\affiliation{ 
Department of Polymer Science and Engineering, University of Massachusetts, Amherst, MA 01003
}%

\renewcommand\thefigure{S\arabic{figure}}
\renewcommand\thesection{S\arabic{section}}
\renewcommand\theequation{S\arabic{equation}}

\maketitle

\section{Strong segregation theory formalism}
\label{sec: SST}

Here, we review the assumptions and necessary details associated with the strong segregation theory (SST) describing thermodynamic equilibria of self-assembled diblock copolymers~\cite{Semenov1985}. In the $\chi N \rightarrow \infty$ limit, chains tend to stretch strongly from the IMDS to reduce the (lateral) area per chain exposure to unlike blocks. 
Since chain repulsion dominates over configurational entropy in this regime, the dominant configuration of the melt has chain conformations that follow ``classical trajectories''~\cite{Milner1988} which minimize the stretching free energy of chains subject to constraints of incompressibilty in the brush.  As a result, blocks of each type are expected follow straight trajectories that extend from the free ends in the bulk of A and B brush to the IMDS, where the junction points are localized. In the $\chi N \to \infty$ limit, the effect of brush interpenetration is subdominant, occurring over a distance that vanishes relative to equilibrium brush height~\cite{Goveas1997}.   Hence, capturing the leading free energy terms in SST requires us only to consider brush domains that extend up to the ``terminal boundary'' that divides the two contacting brushes in a A or B domain~\cite{Reddy2021}. Additionally, in the strong segregation limit (SSL), the width of the interface between A and B blocks vanishes relative to brush height, allowing us to approximate it as a 2D surface to which junction points between A and B are localized.

The {\it total} free energy ${\cal F}$ (in units of $k_{\mathrm{B}}T$) of the system of polymer chains in melt has two components
\begin{equation}
{\cal F} = {\cal F}_{\rm int} + {\cal F}_{\rm st}
\end{equation}
where ${\cal F}_{\rm int}$ represents the enthalpic contribution due to interblock repulsion at AB interface and ${\cal F}_{\rm st}={\cal F}^{\rm A}_{\rm st}+{\cal F}^{\rm B}_{\rm st}$ is the entropic cost of chain stretching, decomposed into contributions from A and B molten brush domains~\cite{Olmsted1998, Likhtman1994}. 

In this study, we compute the stretching energy of chains using two approaches: the \textit{parabolic brush theory} (PBT)~\cite{Milner1988} and its {\it end-exclusion zone corrected} (EEZC)  form~\cite{Ball1991, Belyi2004}.  In the PBT, the free-energy minimizing balance of chain entropy and incompressibilty results in a free energy density that is proportional to the second moment of the volume relative to the IMDS~\cite{Matsen2002}. The stretching free energy contribution from the $\alpha = {\rm A, B}$ block has the form 
\begin{equation}\label{eq:PBT}
{\cal F}_{\rm st}^{\alpha} = \frac{3 \pi^2 \rho}{8 N^2_{\alpha} a_{\alpha}^{2}} \int_{V_{\alpha}} \mathrm{d}V ~z^{2}
\end{equation}
where $z$ is the extension of a chain away from the IMDS along a straight line, as shown in Fig.~2A in the main text.
Here, $\rho^{-1}$ is the volume occupied by single monomer unit, $N_\alpha$ is the number of segments in block $\alpha$ and $a_{\alpha}$ is the corresponding statistical segment length. The volume integral is taken over volume $V_{\alpha}$ of the brush domains of type $\alpha$. 

The PBT makes implicit assumptions about the free end distribution $g(z)$ throughout incompressible brushes that results in a well-documented error~\cite{Ball1991} from brushes that splay away from grafting surfaces, usually associated with positive mean curvature of those surfaces, e.g.~the outer domains of cylindrical or spherical domains.  In those cases, the PBT requires an unphysical negative free end distribution~\cite{Semenov1985}, indicating that this theory is strictly only an upper-bound on the true SSL limit for those local regions.  The exact SSL requires the introduction of an end-exclusion zone (EEZ) ~\cite{Ball1991}, for which $g(z) =0$, near to the anchoring surface (i.e.~the IMDS), and leads to a more complex, self-consistent distribution in the brush, resulting in a free energy density that deviates at least somewhat from the simple $\propto z^2$ from of PBT, i.e. eq. (\ref{eq:PBT}).  To capture and assess the importance of free energy corrections from the EEZ to the PBT as summarized in Sec.~\ref{sec: eez}, we implement a variant of the approach of ref.~\cite{Belyi2004}, described in \cite{Dimitriyev2021Arxiv}, to capture the exact SSL entropy for molten brush domains of arbitrary local geometries, of the type needed to construct complex DG domains.  Notably, for the purposes of comparing the free energies of Lam, Hex and DG near to their phase boundaries, we find that the PBT is sufficient to an excellent approximation.  Moreover, given the intuitive geometrical form of PBT entropy (i.e.~second-moment of brush volumes relative to the IMDS) we focus most of the main text presentation on its simpler form, and results derived from it.

To determine the optimal free of a given morphology $X$, we evaluate the free energy per chain $F(x)$ based on a particular periodic repeat unit of the structure (e.g.~a unit cell, or an asymmetric unit), which has a volume $V(x)$, corresponding to $\rho V(X)/N$ chains per unit, where $N=N_{\rm A} +N_{\rm B} $ is the total chain length.  Following the definitions of IMDS area $A(X)$ and second-moments $I_{\alpha}(X)$ in these repeat units and dividing by the number of chains per unit, we have the form of eq.~(1) in the main text, where the surface tension of the IMDS is
\begin{equation}
\gamma = \rho \bar{a} \sqrt{\frac{\chi}{6}} \bigg( \frac{2}{3} \frac{\epsilon_0 ^{2/3} - \epsilon_0 ^{-3/2}}{\epsilon_0 - \epsilon_0 ^{-1}} \bigg)
\end{equation}
where $\bar{a} \equiv \sqrt{a_{\rm A}a_{\rm B}}$ is the geometric mean of segment lengths and
$\epsilon_0 \equiv a_{\rm A}/ a_{\rm B}$ is the conformational asymmetry parameter. The coefficients of the stretching free energies, $\kappa_{\alpha}$, are determined by molecular architecture,
\begin{equation}
\begin{split}
    \kappa _{\rm A} = \frac{3 \pi ^{2} \rho \bar{n}^2}{4 N^2 \bar{a}^{2}} \frac{1}{f^{2}\epsilon}    \\
    \kappa _{\rm B} = \frac{3 \pi ^{2} \rho \bar{n}^2}{4 N^2 \bar{a}^{2}} \frac{\epsilon}{(1-f)^{2}},  
\end{split}
\end{equation}
where $f = N_{\rm A}/N$ is the volume fraction of A-segments per chain and $\epsilon \equiv (n_{\rm B} a_{\rm A})/(n_{\rm A} a_{\rm B})=(n_{\rm B}/n_{\rm A}) \epsilon_0$ is the {\it elastic asymmetry} parameter defined by Milner for a generalized class of ${\rm A}_{n_{\rm A}} {\rm B}_{n_{\rm B}}$ miktoarm star block copolymers~\cite{Milner1994}, and $\bar{n} = \sqrt{n_{\rm A} n_{\rm B}}$.

Computing $A(X)$ and $I_{\alpha}(X)$ requires construction of space-filling configurations of local brush regions (with corresponding chain trajectories), divided by an IMDS.  For all geometries besides Lam, these local chain geometries are non-uniform, and we approximate the continuous integrals over the brush volume using tessellations into narrow wedge-like volumes that extend from the IMDS in along straight paths up to the terminal boundaries, with the accuracy of the discrete approximation improving with increasingly narrow wedge volumes.  In Secs.~\ref{sec: medial} and \ref{sec: hex} we describe the respective approaches to construct tessellations for DG, based on the medial construction, and for Hex, based on a ``kinked-path'' ansatz.

\section{Area per chain and equilibrium thermodynamics}

\label{sec: area}
Here, we describe the SST thermodynamics in terms of the {\it IMDS area per chain} $\Sigma$. $\Sigma$ varies thermodynamically, with an optimal value that is determined by the competition between IMDS enthalpy and change stretching, which also sets the equilibrium domain size. Additionally, this allows us to compare the free energy of distinct morphologies $X$ by holding them at a constant value $\Sigma$, a state at which their free energies derive only from relative differences in stretching energy of block $\alpha$ (either A or B). Equal IMDS area per chain comparisons are shown in main text Figs.~3,4. 

Given a system with a reference length scale $D_0$ (which might correspond e.g.~to a the length of translation vector in a periodic structure) and reference mean area per chain $\Sigma_0$, the area per chain $\Sigma$ is related by an appropriate affine rescaling of the system $L_0 \mapsto L = \lambda L_0$.  This affine rescaling alters the geometric quantities that enter the free energy as follows
\begin{align}
\label{Equalarea2}
\begin{split}
V_0 &\mapsto V = \lambda^{3} V_{0} \\
A_0 &\mapsto A =  \lambda^{2} A_{0} \\
I_{\alpha,0} &\mapsto I_\alpha = \lambda^{5} I_{\alpha ,0} 
\end{split}\,\,,
\end{align}
which allows us to express the area per chain $\Sigma$ using scale factor $\lambda$
\begin{equation}
    \Sigma = \frac{A_{0}}{V_{0}} \frac{N}{\rho} \frac{1}{\lambda} \,\, .
\end{equation}
Alternatively this gives the length scaling needed to set the average area per chain to a given value $\Sigma$
\begin{equation}
    \lambda = \frac{A_{0}}{V_{0}} \frac{N}{\rho} \frac{1}{\Sigma} \,\, .
\end{equation}
From this, we have the free energy of morphology $X$ as a function of area per chain,
\begin{equation}
\label{Equalarea4}
    F(X, \Sigma) = \gamma \Sigma + \frac{N^3}{2 \rho^3 \Sigma^2}\big[ \kappa_{\rm A}\mathcal{S}_{\rm A}(X) + \kappa_{\rm B}\mathcal{S}_{\rm B}(X) \big] \,\, ,
\end{equation}
where rescaled dimensionless entropic cost $S_{\alpha}$ of stretching for each block is defined as
\begin{equation}
\label{Equalarea5}
    \mathcal{S}_{\alpha} = \frac{I_{\alpha,0}}{V_{0}} \bigg(\frac{A_{0}}{V_{0}}\bigg)^{2} \, .
\end{equation}
Thus, at fixed mean area per chain $\Sigma$, the differences in free energy of competing phases is completely determined by the quantity in the brackets $[\cdots]$ of eq.~(\ref{Equalarea4}), viz.~$S_{\rm A}(X)$ and $S_{\rm B}(X)$ while the coefficients $\kappa _{\rm A}$ and $\kappa _{\rm B}$ are same for all phases at any given molecular parameters.

Finally, we consider the equilibrium domain size by minimizing $F(X, \Sigma)$ with respect to $\Sigma$, resulting in a optimal area per chain,
\begin{equation}
    \Sigma_*(X) = \frac{N}{\gamma^{1/3} \rho}\big[ \kappa_{\rm A}\mathcal{S}_{\rm A}(X) + \kappa_{\rm B}\mathcal{S}_{\rm B}(X) \big]^{1/3}
\end{equation}
and equilibrium free energy per chain
\begin{equation}
\label{eq: Fstar}
    F_*(X) = \frac{3\gamma^{2/3} N^{1/3} \bar{n}^{2/3}}{2 \rho^{2/3} a^{2/3}}\big[ \bar{\kappa}_{\rm A}\mathcal{S}_{\rm A}(X) + \bar{\kappa}_{\rm B}\mathcal{S}_{\rm B}(X) \big]^{1/3}
\end{equation}
where $\bar{\kappa}_\alpha = N^2 \bar{a}^2 \rho^{-1} \bar{n}^{-2} \kappa_\alpha $ is the dimensionless block stiffness coefficient, functions of only on $f$ and $\epsilon$

\section{Medial tessellations of TPN for SST}

\label{sec: medial}

We use the medial map to generate space filling tessellations of triply periodic network morphologies.  Here we focus on an example for the DG morphology, but this method generalizes to morphologies of arbitrary shape and topology.

The process begins with a set of orientated generating surfaces \textbf{G}, where the orientation corresponds to directions pointing toward the ``inner'' and ``outer'' domains.  
For the case of the DG network, as we detail below, \textbf{G} can, for example, be supplied by a level set $F({\bf x}) = 0$ of a function $F$ possessing $I4_{1}32$ symmetry, e.g.
\begin{equation}
\label{gensurf1}
F(x,y,z) = \sin \bigg[\frac{2 \pi  x}{D}\bigg] \cos \bigg[\frac{2 \pi  y}{D}\bigg] + \cos \bigg[\frac{2 \pi  x}{D}\bigg] \sin \bigg[\frac{2 \pi  z}{D}\bigg]+\sin \bigg[\frac{2 \pi  y}{D}\bigg] \cos \bigg[\frac{2 \pi  z}{D}\bigg]-\phi_0
\end{equation}
where $D$ is the length of the DG unit cell and $\phi_0$ is the variable level set parameter.
For $\phi_0 = 0$, this is an approximation of the gyroid G minimal surface. 
For $\phi_0 \neq 0$, solutions to eq.~(\ref{gensurf1}) result in enantiomeric pairs of surfaces ($\phi_0$ and $-\phi_0$) that have the same topology as the IMDS separating tubular domains and matrix domain of DG.
This level set equation is generalized in a later Sec.~\ref{sec: parametric} below.

Given a set of generating surfaces, we construct a triangular mesh to use as a discrete approximation to this surfaces; to triangulate such implicit surfaces, we use the \texttt{CGAL} C++ library \cite{CGAL}.  We then construct the medial sets of this generating surface \textbf{G}, following a method described in the SI ref.~\cite{Reddy2021}. Briefly, this is done following the algorithm of ref.~\cite{Schroder2003} by finding the Voronoi cells of each vertex of the mesh approximation to \textbf{G}, which is done using the \texttt{Voro}++ software \cite{Voro}.  The medial sets are then generated by finding the intersection of the normal to the generating surface with the outer boundary of the point's Voronoi cell at its two distant ends (on both sides of the surface), representing locations that are equidistant to multiple points on the boundary mesh.  That pair of points are ``images'' of generating vertex on \textbf{G} mapped to the medial sets on either side of the generating surface (for DG these correspond to tubular and matrix domains). 
As a convention for the remainder of this SI, we denote vertices on the medial set of the matrix with a ``-'' and vertices in the tubular domains with a ``+''.   We denote $\hat{\mathbf{n}}_{\pm}$ as the medial vector pointing normally from the vertex at \textbf{G} to its medial image inside the tubular and matrix domains for ``+'' and  ``-'', respectively. 
The result is a medial map: for each vertex $\mathbf{v} \in \mathbf{G}$, there is a pair of points $\mathbf{v}_{\rm -} \in \mathbf{T}_{\rm B}$ (in the  matrix domain) and $\mathbf{v}_{\rm +} \in \mathbf{T}_{\rm A}$ (in the tubular domain) given by the map ${\bf m}_\alpha$,
\begin{equation}\begin{split}
\label{medialmap}
{\bf m}_\alpha\mkern+8mu:\mkern+8mu \mathbf{G} &\to \mathbf{T}_\alpha \\
\mathbf{v} &\mapsto \mathbf{v}_\alpha = \mathbf{v}+d_{\alpha}(\mathbf{v})\hat{\mathbf{n}}_{\alpha}(\mathbf{v}) = {\bf m}_\alpha(\mathbf{v})
\end{split}\end{equation}
where $d_{\alpha}(\mathbf{v})$ is the distance between $\mathbf{v}$ and $m_\alpha(\mathbf{v})$ found from the medial set construction.  Notice that at this point we identify the medial sets $\mathbf{T}_{\rm A}$ and $\mathbf{T}_{\rm B}$ as {\it terminal boundaries}, although {\bf G} should not be confused with an IMDS as the structure does not necessarily satisfy local balance of volume fractions.

The medial map allows us to decompose space into a set of volumes, each of which is associated to an initial facet, which we label $\mu$, of the initial generating surface mesh. By construction, every mesh facet $(\mathbf{v}^{(1)}_{\mu},\mathbf{v}^{(2)}_{\mu},\mathbf{v}^{(3)}_{\mu})$ on $\mathbf{G}$ is mapped to a corresponding facet $ (\mathbf{v}^{(1)}_{\alpha,\mu},\mathbf{v}^{(2)}_{\alpha,\mu},\mathbf{v}^{(3)}_{\alpha,\mu})$ on $\mathbf{T}_\alpha$, with the connectivity of neighbor edges $\mathbf{T}_\alpha$ inherited from the connectivity of corresponding vertices on {\bf G}.
Since this map occurs via straight lines connecting the vertices of the facets (spanning from ``inner'' to ``outer'' medial set), as shown in eq.~(\ref{medialmap}), we can construct a space-filling tessellation of the region between $\mathbf{T}_{\rm A}$ and $\mathbf{T}_{\rm B}$.
This polyhedral region spans between associated  facets on these medial sets defined by the points $\{\mathbf{v}_{\rm A}^{(1)},\mathbf{v}_{\rm A}^{(2)},\mathbf{v}_{\rm A}^{(3)}\}$ in $\mathbf{T}_{\rm A}$, forming one triangular face and $\{\mathbf{v}_{\rm B}^{(1)},\mathbf{v}_{\rm B}^{(2)},\mathbf{v}_{\rm B}^{(3)}\}$ in $\mathbf{T}_{\rm B}$, forming a second triangular face.
Pairs of identified vertices $\{(\mathbf{v}_{\rm A}^{(1)},\mathbf{v}_{\rm B}^{(1)}),(\mathbf{v}_{\rm A}^{(2)},\mathbf{v}_{\rm B}^{(2)}),(\mathbf{v}_{\rm A}^{(3)},\mathbf{v}_{\rm B}^{(3)})\}$ form three edges connecting the triangular faces, adding three quadrilateral faces to the resulting polyhedron, resulting in a 5-face polyhedron, or pentahedron.
Note that the quadrilateral faces may be non-planar in general, so the resulting shape is a curved pentahedron, which we shall refer to as a ``wedge.''  Next we describe a simple map that spans the interior of these wedges such that quadrilateral faces of the wedges are ruled surfaces that perfectly match without gaps from wedge to neighbor wedge, and therefore the collection of triangular prismatic wedges on either side of {\bf G} is space filling  between medial sets $\mathbf{T}_{\rm A}$ and $\mathbf{T}_{\rm B}$.

\begin{figure}
\center
\includegraphics[width=1\textwidth]{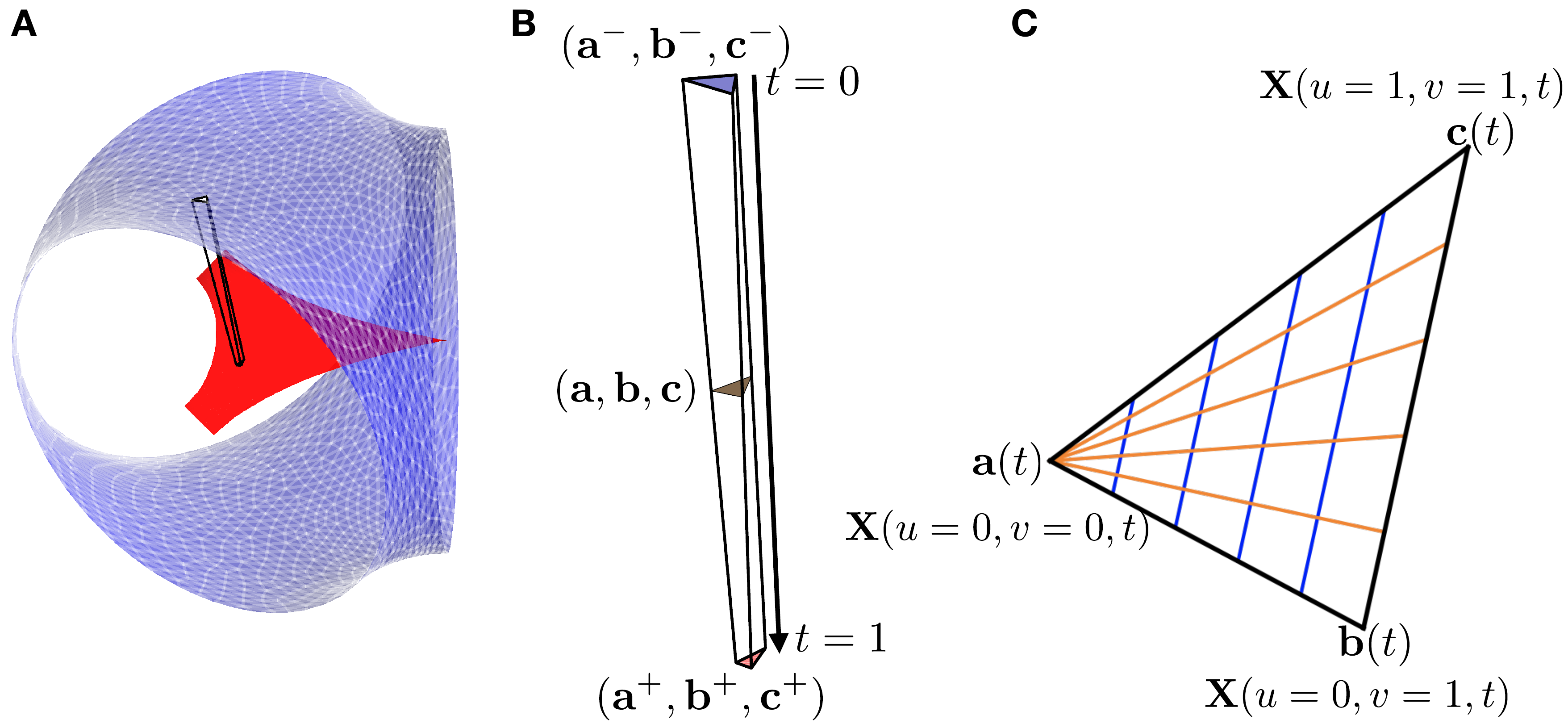}
\caption{\label{fig:wedge_geometry} (A) Illustration of the tessellation of a nodal region, bounded by \textbf{T}\textsubscript{A} and \textbf{T}\textsubscript{B}, by prismatic wedges. (B) A wedge is capped by faces $(\mathbf{a}^-,\mathbf{b}^-,\mathbf{c}^-)$ on \textbf{T}\textsubscript{B} and $(\mathbf{a}^+,\mathbf{b}^+,\mathbf{c}^+)$ on \textbf{T}\textsubscript{A}. Sections of the wedge are parameterized by coordinate $t\in[0,1]$. (C) Triangular section of the wedge at fixed $t$, parameterized by coordinates $u,v \in[0,1]$, with lines of constant $u$ (orange) and $v$ (blue). Any point $\mathbf{X}$ in the wedge can be uniquely specified by $(u,v,t)$.}
\end{figure}

We parameterize the interior of volume-tessellating wedges by considering a mapping between 2D triangular sections of the wedges -- parameterized by coordinates $(u,v)$ -- and coordinate $t$ that translates along the length of the wedge to all points $\mathbf{X}$ within the wedge interior. To construct the coordinates we consider a given wedge with associated connected triangular facets on $\mathbf{T}_{\rm B}$ and $\mathbf{T}_{\rm A}$, which we denote as the ``-'' and ``+'' faces of the wedge.  The vertices on those respective faces are denoted by $\{\mathbf{a}_-,\mathbf{b}_-,\mathbf{c}_-\}$ and $\{\mathbf{a}_+,\mathbf{b}_+,\mathbf{c}_+\}$ such that each pair of vertices $\mathbf{a}_\pm$, $\mathbf{b}_{\pm}$, and $\mathbf{c}_\pm$ are associated by the medial map of the generating surface.
The edge vectors connecting $\mathbf{T}_{\rm B}$ to $\mathbf{T}_{\rm A}$ are given by 
\begin{equation}\begin{split}
    \mathbf{C}_a &= \mathbf{a}_+ - \mathbf{a}_- \\
    \mathbf{C}_b &= \mathbf{b}_+ - \mathbf{b}_- \\
    \mathbf{C}_c &= \mathbf{c}_+ - \mathbf{c}_-
\end{split}\end{equation}
where $\mathbf{C}_a = \mathbf{a}_+ - \mathbf{a}_-$, $\mathbf{C}_b = \mathbf{b}_+ - \mathbf{b}_-$, and $\mathbf{C}_c = \mathbf{c}_+ - \mathbf{c}_-$.
Given these edge vectors, we define a linear map along the three edges $\mathbf{a}(t)$, $\mathbf{b}(t)$, and $\mathbf{c}(t)$ 
\begin{equation}\begin{split}
    \mathbf{a}(t) &= \mathbf{a}_- + \mathbf{C}_a t \\
    \mathbf{b}(t) &= \mathbf{b}_- + \mathbf{C}_b t \\
    \mathbf{c}(t) &= \mathbf{c}_- + \mathbf{C}_c t
\end{split}\end{equation}
where $t \in [0,1]$ denotes the length fraction along each edge such that $\mathbf{a}(0) = \mathbf{a}_-$, $\mathbf{a}(1) = \mathbf{a}_+$, $\mathbf{b}(0) = \mathbf{b}_-$, $\mathbf{b}(1) = \mathbf{b}_+$, $\mathbf{c}(0) = \mathbf{c}_-$, and $\mathbf{c}(1) = \mathbf{c}_+$.
At given $t$, these three vertices define a triangular region --- defined by $\{\mathbf{a}(t),\mathbf{b}(t),\mathbf{c}(t)\}$ --- that continuously interpolates between the upper facet on $\mathbf{T}_{\rm B}$ at $t=0$ and bottom facet on $\mathbf{T}_{\rm A}$ at $t=1$ .
Taking advantage of the convexity of each triangular region at fixed $t$, any point $\mathbf{X}$ within a triangle can be described by a pair of coordinates $(u,v) \in [0,1]\times[0,1]$, via
\begin{equation}\label{eq:wedge_param}
\mathbf{X}(u,v,t) = (1-v)\mathbf{a}(t) + v(1-u)\mathbf{b}(t) + uv\mathbf{c}(t)    
\end{equation}
giving us a coordinate system $(u,v,t)$ for any point within a wedge (i.e.~sweeping each coordinate from 0 to 1 spans the full volume interior to the wedge).  Careful consideration shows that quadrilateral (long) faces of wedges are ruled surfaces, where the edge vectors at constant $t$ are the rulings.  Moreover, these rulings are identical for face sharing neighbor wedges, since the linear map with $t$ from - to + vertices is the same for both wedges.  Hence, and crucially, this ensures that wedge faces match perfectly without gaps.  

Using eq.~(\ref{eq:wedge_param}), the differential volume element is given by
\begin{equation}
\mathrm{d}V = \mathrm{d}u~\mathrm{d}v~\mathrm{d}t \left[\frac{\partial \mathbf{X}}{\partial t} \cdot \bigg( \frac{\partial \mathbf{X}}{\partial u} \times \frac{\partial \mathbf{X}}{\partial v} \bigg)\right]
\end{equation}
where the $J(u,v,t)=\left[\frac{\partial \mathbf{X}}{\partial t} \cdot \bigg( \frac{\partial \mathbf{X}}{\partial u} \times \frac{\partial \mathbf{X}}{\partial v} \bigg)\right]$  is the determinant of the Jacobian of the coordinate mapping $(u,v,t)\mapsto \mathbf{X}$.
To evaluate the Jacobian, we note that the partial derivatives of $\mathbf{X}$ are given by
\begin{equation}\begin{split}
    \frac{\partial \mathbf{X}}{\partial u} &= v \mathbf{e}_{bc}(t) \\
    \frac{\partial \mathbf{X}}{\partial v} &= \mathbf{e}_{ba}(t) + v \mathbf{e}_{bc}(t)  \\
    \frac{\partial \mathbf{X}}{\partial t} &= (1-v)\mathbf{C}_a + v(1-u)\mathbf{C}_b + uv\mathbf{C}_c
\end{split}\end{equation}
where $\mathbf{e}_{ba}(t) \equiv \mathbf{a}(t) - \mathbf{b}(t)$ and $\mathbf{e}_{bc}(t) \equiv \mathbf{c}(t) - \mathbf{b}(t)$ are the edge vectors of the triangle formed at fixed $t$.  Note that map from $(u,v,t)$ to $\mathbf{X}$ is one-to-one for all points in the wedge provided that $J(u,v,t)$ is positive definite, which corresponds to the condition that the interior facet area is non-vanishing at all $t\in[0,1]$.    
Integrating over cross-section coordinates $u$ and $v$, we find an expression for the amount of volume ${\rm d}V(t)$ per edge fraction interval ${\rm d}t$, given by
\begin{equation}\begin{split}
    \frac{\partial V(t)}{\partial t} &= \iint_0^1 {\rm d}u\,{\rm d}v\, \left[\frac{\partial \mathbf{X}}{\partial t} \cdot \bigg( \frac{\partial \mathbf{X}}{\partial u} \times \frac{\partial \mathbf{X}}{\partial v} \bigg)\right] \\
    &= \frac{\mathbf{C}}{2}\cdot\left[\mathbf{e}_{bc}(t)\times\mathbf{e}_{ba}(t)\right]
\end{split}\end{equation}
where the centroidal vector $\mathbf{C} \equiv (\mathbf{C}_a+\mathbf{C}_b+\mathbf{C}_c)/3$ joins the centroid of triangle facet at ${\bf T}_{\rm A}$ to the centroid of the facet at ${\bf T}_{\rm B}$ .
The length of $\mathbf{C}$ provides a measure of wedge height, so we define $h \equiv |\mathbf{C}|$.
Expanding $\partial V/\partial t$ in powers of $t$, we find
\begin{equation}
   \frac{\partial V(t)}{\partial t} = \mathbf{C}\cdot \mathbf{A}_-\left[1 + \frac{\mathbf{C}\cdot(\bm{\beta}_1 + \bm{\beta}_2)}{2 \mathbf{C}\cdot\mathbf{A}_-}t + \frac{\mathbf{C}\cdot\bm{\gamma}}{2\mathbf{C}\cdot\mathbf{A}_-}t^2 \right] \, ,
\end{equation}
where $\mathbf{A}_- \equiv \mathbf{e}_{bc}(0)\times\mathbf{e}_{ba}(0)/2$ is the oriented area of the trianglar facet at $t=0$, and $\bm{\beta}_1$, $\bm{\beta}_2$ and $\bm{\gamma}$ are vectors describing the splay geometry of the three connecting edges and are given by
\begin{equation}\begin{split}
    \bm{\beta}_1 &\equiv (\mathbf{C}_c-\mathbf{C}_b)\times \mathbf{e}_{ba}(0) \\
    \bm{\beta}_2 &\equiv  \mathbf{e}_{bc}(0)\times(\mathbf{C}_a-\mathbf{C}_b) \\
    \bm{\gamma} &\equiv (\mathbf{C}_c\times\mathbf{C}_a) + (\mathbf{C}_b\times\mathbf{C}_c) + (\mathbf{C}_a\times\mathbf{C}_b)
\end{split}\end{equation}
The splay of the three edge vectors controls the rate of change of volume with the parameter $t$ in a manner that captures the equivalent Steiner's law describing the area $A_\mu(z)$ of surfaces parallel to a curved surface at $z=0$ in terms of mean curvature $H$ and Gaussian curvature $K_G$ via $A(z) = A(0)(1 + 2Hz + K_Gz^2)$ \cite{Hyde1997_ch4}.
Identifying each power of $t$ with the appropriate dimensionless ``wedge curvatures,'' we have the wedge version of Steiner's law,
\begin{equation}
    A_\mu (z) = \frac{\partial V}{\partial z} = \hat{\mathbf{C}}\cdot \mathbf{A}_-(1 + 2\tilde{H}z + \tilde{K}z^2) \,,
    \label{eq: Az}
\end{equation}
where height along the wedge (measured from ${\bf T}_{\rm A}$) can be defined as $z\equiv h t$ and the wedge analogues of mean and Gaussian curvatures for the wedge are given by
\begin{equation}\begin{split}
    \tilde{H} &\equiv \frac{\mathbf{C}\cdot(\bm{\beta}_1 + \bm{\beta}_2)}{4 h \mathbf{C}\cdot\bf{A}_-}\\
    \tilde{K} &\equiv \frac{\mathbf{C}\cdot\bm{\gamma}}{2 h^2 \mathbf{C}\cdot\bf{A}_-}
\end{split}\end{equation}
respectively.  Here we use the notation $\tilde{H}$ and $\tilde{K}$ to clarify that these do not correspond the curvatures of the IMDS, or the terminal boundaries, as the wedges are in general tilt with respect to the normals of those surfaces

Given this area distribution, it is straightforward to compute the volume of wedge within height $h$,
\begin{equation}
    V_\mu (z) = \int_0^z {\rm d}y \,A_\mu(y) = \hat{\mathbf{C}}\cdot {\bf A}_- z\left(1 + \tilde{H}z + \frac{1}{3}\tilde{K}z^2\right)
    \label{eq: Vz}
\end{equation}
On either side of the IMDS, the wedges enclose polymer blocks with junctions bound to a common point, leading to the condition of {\it volume balance} in each wedge.  Specifically, the region spanning from the IMDS to ${\bf T}_{\rm B}$ must enclose a fraction $1-f$ of the total volume of the wedge (and the corresponding fraction on the A-side encloses $f$).  With this in mind, we determine the balanced height $h_{\rm b}$ of the IMDS within each wedge according to the condition
\begin{equation}\label{eq: divideplane}
    V_\mu (h_{\rm b}) = (1 - f)V_\mu (h) .
\end{equation}
where $z=h$ is the total wedge height
Defining the height fraction of the dividing interface $t_{\rm b} = h_{\rm b}/h$, the volume balance condition is determined by the solution of a cubic equation,
\begin{equation}\label{eq: divideplane2}
t_{\rm b} ^3 \frac{\Tilde{K}h^2}{3} + t_{\rm b} ^2 \Tilde{H}h + t_{\rm b} - (1-f)(1+\Tilde{K}h^2/3+\Tilde{H}h) = 0 .
\end{equation}
We solve this cubic equation for the real root $0<t_{\rm b}<1$ in each wedge to set the location of the intersection of the IMDS with the wedges.

Given this balanced IMDS location in each wedge it is straightforward to compute the second moments of volume relative to the IMDS for each wedge $\mu$,
\begin{equation}
    I_{\mu, {\rm A}} = \int_{h_{\rm b}}^{h} {\rm d}z\,  \hat{\mathbf{C}}\cdot \mathbf{A}_- \big(1 + 2\tilde{H}z + \tilde{K}z^2\big) \big(z - h_{\rm b})^2 =\frac{\hat{\mathbf{C}}\cdot {\bf A}_- (h-h_{\rm b})^3}{3}\Big[1+\frac{\tilde{H}}{2} (h_{\rm b} + 3 h) + \frac{\tilde{K}}{10}(h_{\rm b}^2+3 h_{\rm b} h +6 h^2) \Big]
\end{equation}
and 
\begin{equation}
    I_{\mu, {\rm B}} = \int_{0}^{h_{\rm b}} {\rm d}z\, \hat{\mathbf{C}}\cdot {\bf A}_- \big(1 + 2\tilde{H}z + \tilde{K}z^2\big) \big( h_{\rm b}-z)^2 = \frac{\hat{\mathbf{C}}\cdot {\bf A}_-}{3} \Big[ h_{\rm b}^3 +\frac{1}{2} \tilde{H}h_{\rm b}^4+\frac{1}{10} \tilde{K}h_{\rm b}^5\Big] .
\end{equation}
The total stretching moments per cell computed via simple sum over the wedges within the repeat unit volume
\begin{equation}
    I_{\alpha} =\sum_{\mu \in V_0}I_{\mu, \alpha}.
\end{equation}

\begin{figure}
\center
\includegraphics[width=0.5\textwidth]{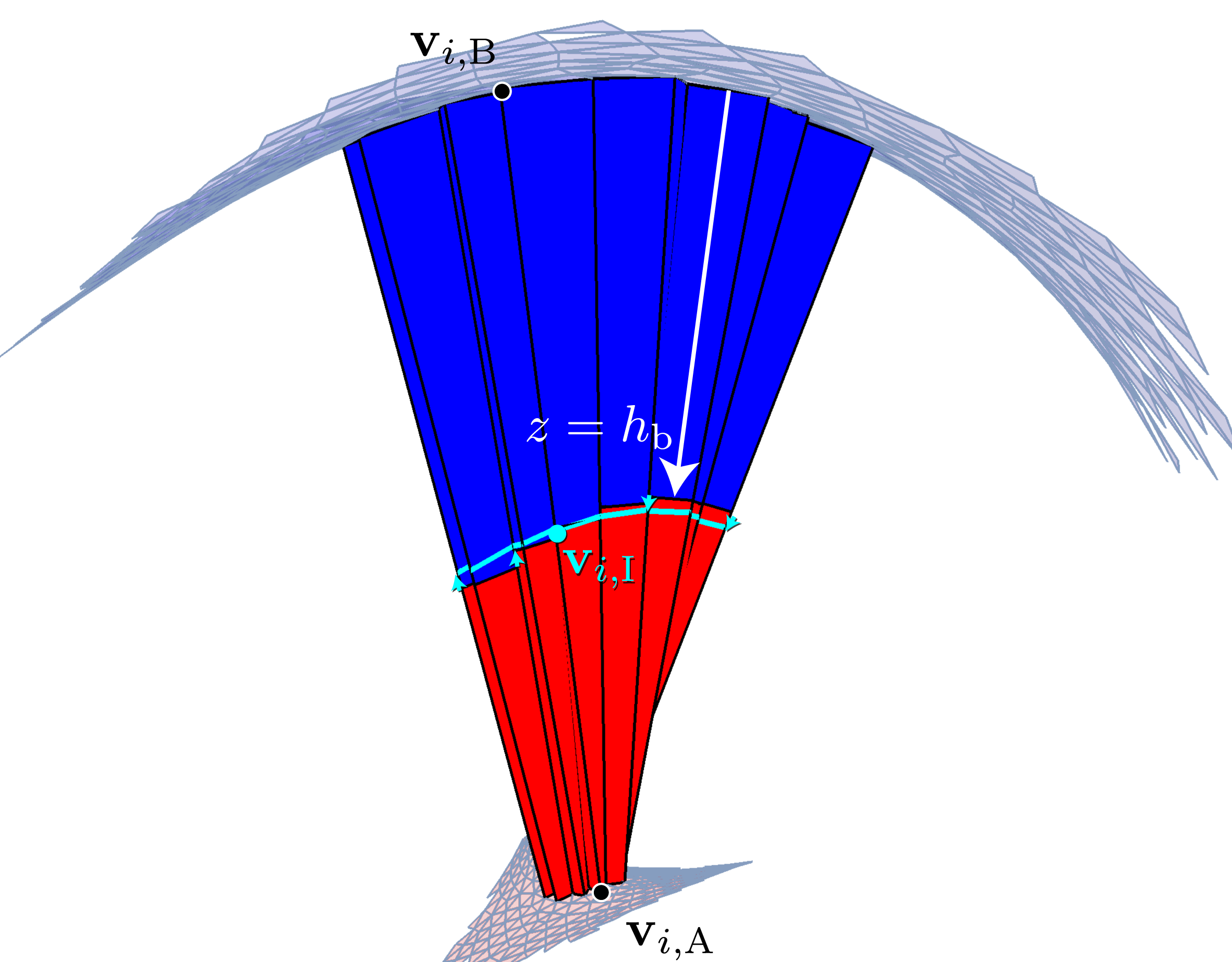}
\caption{\label{fig:wedge_balance} Schematic illustrating the joining of balanced wedges to form a continuous, balanced IMDS. The fraction of red to blue in each wedge is determined by $f$, with a dividing plane at $z = h_{\rm b}$. The dividing planes are then joined together such that neighboring facets meet at vertices $\mathbf{v}_{i,{\rm I}}$ to form a continuous IMDS (cyan).}
\end{figure}

Solving for the dividing plane in each of the wedges ensures that volumes of A and B blocks occupy volume on either side of IMDS at the proper ratio.  However, since the positions of the dividing planes are determined for each wedge independently, the surface formed from the collection of dividing planes is generally not continuous, as depicted in Fig.~\ref{fig:wedge_balance}.  To determine a reasonably smooth model of the IMDS we determine the location of the IMDS ${\bf v}_{i,{\rm I}}$ along the edge vector that connects ${\bf v}_{i,{\rm B}}$ to 
${\bf v}_{i,{\rm B}}$ as follows.  Denote the set of wedges $\mu$ (corresponding to facets) that share the vertex $i$ as $\mathcal{N}_i$.  The fractional distance of the IMDS is simply the average of the vertex positions for the volume balance facets from wedges that share that common vertex.  That is,
\begin{equation}
    {\bf v}_{i,{\rm I}}= {\bf v}_{i,{\rm B}} + \langle t_{i} \rangle \big ({\bf v}_{i,{\rm A}}-{\bf v}_{i,{\rm B}} \big),
\end{equation}
where
\begin{equation}
    \langle t_{i} \rangle = \frac{1}{|\mathcal{N}_i|}\sum_{\mu \in \mathcal{N}_i} t_{\rm b} (\mu)
\end{equation}
where $t_{\rm b} (\mu)$ is balanced fractional height in wedge $\mu$.  From this construction, the IMDS is approximated by a triangular mesh whose positions closely follow the locations of balanced facets within the wedge, but with different orientations (see Fig.~\ref{fig:wedge_geometry}) below.  This construction only \emph{slightly} alters the local volume balance in exchange for ensuring a continuous mesh approximation to the IMDS. 


Given the area elements of this IMDS mesh $A_{\mu, {\rm I}}$, the IMDS area used in the mSST calculations follows from simple sum within the repeat volume
\begin{equation}
A_0= \sum_{\mu \in V_0} A_{\mu, {\rm I}} .
\end{equation}




\begin{figure}
\center
\includegraphics[width=1\textwidth]{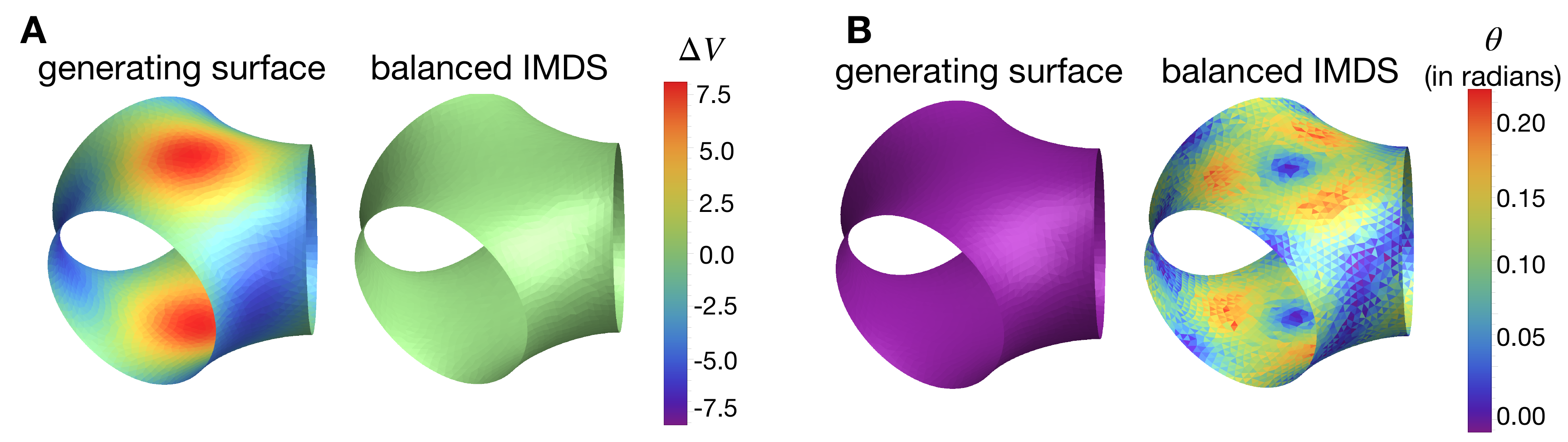}
\caption{\label{fig:wedge_geometry} (A) Comparison of deviation from local volume balance for a generating surface (left) and a balanced IMDS (right). While the generating surface may balance the total volume, only the balanced IMDS achieves local balance. (B) Comparison of tilt between surface normals and wedge orientation for a generating surface (left) and a balanced IMDS (right). The medial property ensures that wedges are aligned with the normal of the generating surfaces, which cannot be maintained if the local volume balance constraint is enforced.}
\end{figure}

In Fig.~\ref{fig:wedge_geometry} we illustrate the consequences of volume balance, which leads to differences between the generating surface ${\bf G}$ and the final IMDS (shown respectively on the left and right of panels in the figure).  The figure shows a generating surface (a level set model of a tubular DG surface) whose combined interiors encloses 50\% of the total volume.  Targeting a composition $f=0.5$, this surface is balanced globally (i.e. the tubular and matrix domains each enclose 50\% of the volume) but not locally.  This is shown my mapping the local differences in volume elements $\delta V_{\rm A/B}$ in the wedges on either side of the IMDS, via the quantity
\begin{equation}
    \Delta V = \frac{\delta V_{\rm A}-\delta V_{\rm B}}{\delta V_{\rm A}+\delta V_{\rm B}} ,
\end{equation} 
which is plotted in Fig.~\ref{fig:wedge_geometry}A.  Viewed as two molten brushed sharing a common anchoring surface at the IMDS, the local density of A chains reaching the IMDS at any point must equal to the number reaching it from the B side, since they share a common junction.  Hence, locally unbalanced configurations (i.e.~where $\Delta V \neq 0$) do not correspond to space-filling configurations of AB diblocks, which is the case for the initial generating surface.  More generally, for $f \neq 0$, the local balance condition corresponds to $\delta V_{\rm A}/\delta V_{\rm B}= f/(1-f)$ everywhere.

Performing the volume balance operation as described above results in a uniform $\Delta V$, but as alluded to above leads to a reorientation of the IMDS relative to the orientation of the wedges.  This can be seen by comparing the maps of local tilt shown in Fig.~\ref{fig:wedge_geometry}B for the unbalanced generating surface to the balanced IMDS.  The tilt angle $\theta$ is measured by the relative orientation between the generating surface or IMDS normal ${\bf \hat{N}}$ and the wedge vector ${\bf C}$ that describes the mean trajectory of the chains in the wedge
\begin{equation}
    \theta = \arccos\Big({\bf \hat{N}} \cdot {\bf C} / h\Big).
\end{equation}
Fig.~\ref{fig:wedge_geometry}B shows that the wedges are normal to the generating surface, which follows directly from the medial construction that maps points in the domain onto the closest points on {\bf G}.  In contrast, trajectories in balanced DG are variably tilted away from the IMDS normal.  Hence, we may view the tilt of chains relative to the IMDS as measure of {\it deviation} of the packing from a strictly medial geometry. Below, in Sec.~\ref{sec: tilt} we revisit this morphological feature in the comparison between mSST and SCFT predictions of DG structure.

\begin{figure}
\center
\includegraphics[width=0.8\textwidth]{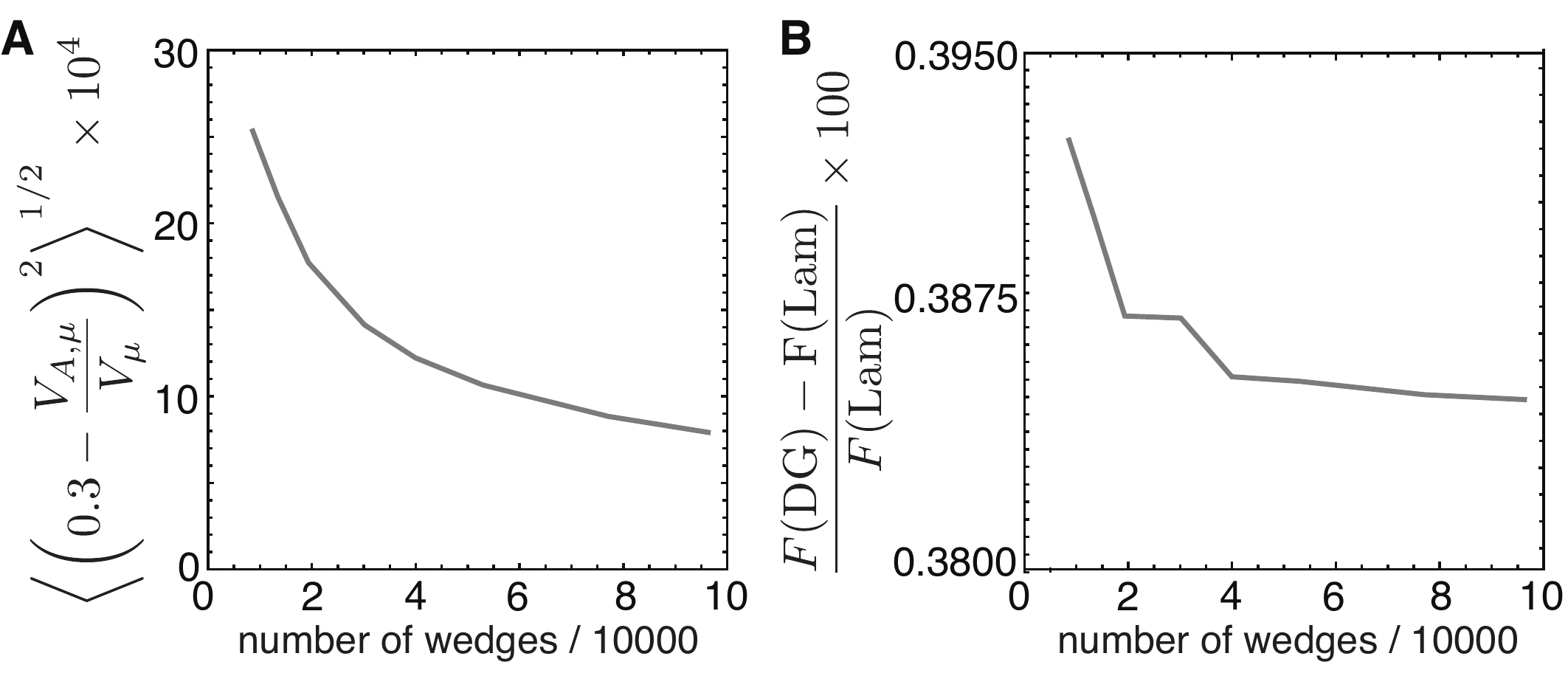}
\caption{\label{fig:number_wedges} (A) Variance of deviation of local volume balance for fixed $\epsilon = 1$, $f=0.3$ upon IMDS adjustment is plotted for varying number of wedges per tessellation and in (B) we plot the relative change to $F(\rm{DG})$ for the same conditions as in (A)}
\end{figure}

Our medial constructions ultimately derive from the finite triangulations of the generating surface, i.e. the number of wedges in the final structure is equal to the number of generating surface facets in the repeat volume.  To minimize the resolution dependence of our mSST construction we consider wedge elements within a single {\it nodal unit} of DG, a 3-valent region as shown in Fig.~1 B-C of the main text, which encloses one of the 16b Wyckoff positions of $Ia\bar{3}d$.  Our final calculations derive from tessellations that include roughly 60,000 wedges per nodal unit.  To assess the error introduced by this discretization, in Fig.~\ref{fig:number_wedges}A we plot the variance of the deviation of the local volume balance for a fixed $f = 0.30$ upon increasing the number of wedges per single nodal unit and show that the error is small and decreases with finer tessellation, which shows that the IMDS adjustment leads to corrections of order less than $10^{-3}$ of the number of wedges used.  We also show the changes to the free energy per chain within a single nodal region for the same $f = 0.30$ by plotting the percentage change relative to $\rm{Lam}$ in Fig.~\ref{fig:number_wedges}B. Beyond the 60,000 number of wedges per tessellation, the changes in fractional free energy are $\lesssim 10^{-3}\%$. As we show in Sec.~\ref{sec: eez} where we consider explicitly the corrections of the end-exclusion zone correction, free energy corrections at this scale (less than the corrections due to end-exclusion) have a negligible effect on predicted phase boundaries.

\section{Parametric variation of generating surfaces}

\label{sec: parametric}

Here we describe the variational framework we use to compute the thermodynamic optimal packing of BCPs according to mSST by sampling over a range of terminal geometries for given TPN symmetry.  In this paper, the focus is on DG structures, but the methodology generalizes to other symmetries.
The approach is to minimize the SST free energy based on balanced medial packings sampling over broad class of generating surfaces \textbf{G} whose symmetries are consistent with the $Ia\bar{3}d$ symmetry of DG morphologies, but differ in shape.
For convenience we parameterize the variable \textbf{G} as level sets the space group $I4_{1}32$, which is a subgroup of $Ia\bar{3}d$.  Specifically, level sets of this lower symmetry are used to generate one enantiomeric tubular single-gyroid (SG) surface, which we then invert and superpose to generate the DG structure. We represent these surfaces as the level set $F(x,y,z)=0$ with Fourier series representation
\begin{equation}
F(x,y,z) = \phi_{1} F_{1,1,0}(x,y,z) + \phi _{2} F_{2,1,1}(x,y,z) + \phi _{3} F_{2,2,0}(x,y,z) + ... -\phi_0
\end{equation}
where basis functions $F_{hkl}(x,y,z)$ are collections of Fourier modes in the $(hkl)$ Bragg plane, where the possible $(hkl)$ planes are determined by the symmetries of the single gyroid $I4_{1}32$, $\phi_i$ are (real) coefficients of the basis functions, and $\phi_0$ is a parameter that shifts the level sets (or equivalently the $(000)$ mode of the Fourier expansion).
Note that the ordering of $\phi_i$ in relation to indices $(hkl)$ is determined by the magnitude of the Bragg vector, $\sqrt{h^2 +k^2 + l^2}$.

For our calculations, it is sufficient to set the periodicity of the reference unit cell to $D_0 = 1$.  In this study, we consider the set of single-gyroid surfaces within the first 3 Fourier modes, the level sets of which are parameterized by $\phi_{0}/\phi_{1}=p_1$ and $\phi_{2}=p_2$ which constitute the 2D variational space of generating surface shapes.  These first two basis functions are
\begin{equation}
    F_{1,1,0}(x,y,z) = \sin (2 \pi  x) \cos (2 \pi  y) + \cos (2 \pi  x) \sin(2 \pi  z)+\sin (2 \pi  y) \cos (2 \pi  z)
\end{equation}
and
\begin{equation}
\begin{split}
    F_{2,1,1}(x,y,z) = \sin (4 \pi  x)\cos (2 \pi  y)\sin (2 \pi  z) + \sin (4 \pi  y)\cos (2 \pi  z)\sin (2 \pi  x) \\ + \sin (4 \pi  z)\cos (2 \pi  x)\sin (2 \pi  y)
\end{split}
\end{equation}
A DG structure of interwoven single gyroids is constructed from the pair of positive and negative level set parameters $\pm \phi_0$ for given set of $\phi_1$ and $\phi_2$ (i.e. $p_i \mapsto -p_i)$.  Notably, volume fractions enclosed in these generating surfaces are {\it not constrained} to be close to ultimate target volume fraction of the mSST morphology.

\begin{figure}
\center
\includegraphics[width=.65\textwidth]{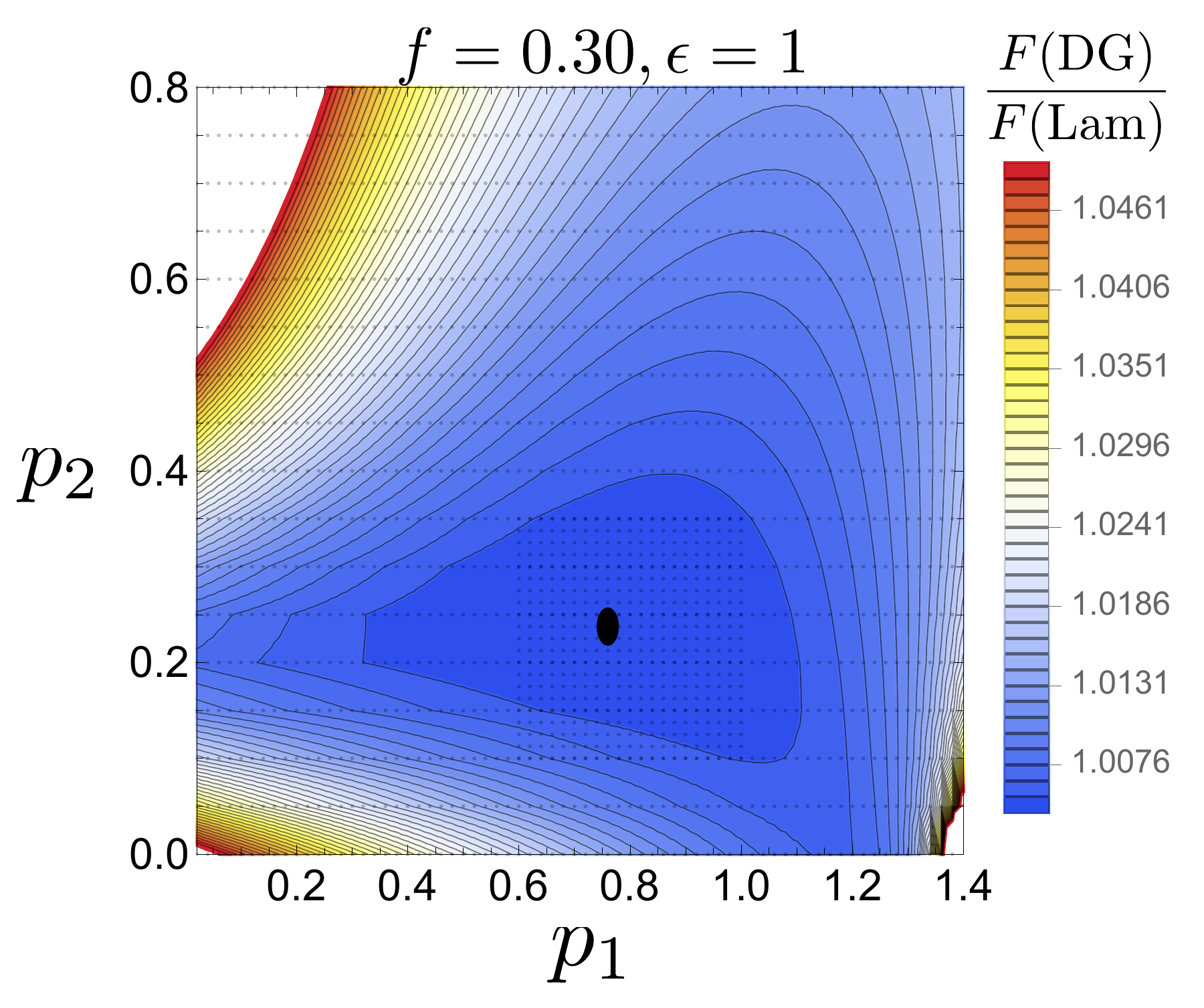}
\caption{\label{fig:free_energy_landscape} Interpolated free energy landscape describing the results of the mSST construction with $f=30$ and $\epsilon = 1$. Each small point drawn over the contour plot represents a single generating surface, parameterized by $p_1$ and $p_2$. To better resolve the location of the free energy minimum, a finer grid of generating surfaces was sampled near the expected free energy minimum (large point).}
\end{figure}

To systematically apply the mSST procedure established in the previous sections (i.e.~find the medial sets, determine the balanced wedge tessellation, and finally compute the free energy), we discretize the space of generating surfaces \textbf{G} into a 2D grid of $p_1$ and $p_2$ parameters. We vary $p_{1}$ in intervals of 0.02 and for $p_{2}$ we start with intervals of 0.05. As described in Sec.~\ref{sec: medial}, the medial sets of a given \textbf{G} are used to generate terminal boundaries of the tubular and matrix domains, $\mathbf{T}_{\rm A}$ and $\mathbf{T}_{\rm B}$, from which we can build a space-filling and locally volume-balanced tessellation by wedges.  Using the IMDS areas and second-moments of volume of these wedges and eq.~(\ref{eq: Fstar}), we compute the free energy of each structure over a discrete 2D parameter grid for a given pair of fixed molecular parameters $(\epsilon,f)$, leading to a free energy density landscape as shown in Fig.~\ref{fig:free_energy_landscape}.  
We ensure that the grid is over a sufficient data range that we can bracket a free energy minimum for a fixed $(f,\epsilon)$. In the vicinity of the minimum of free energy, we further refine the grid by adding grid points in $p_{2}$ with an interval of 0.0125 to determine the minimum free energy via an interpolation in Mathematica.  We find that this procedure ensures that for range of $f$ near the $F(\rm{Lam}) = F(\rm{Hex})$ phase boundary for a particular $\epsilon$, the minimum values of $(p_{1},p_{2})$ smoothly varies with $f$. We illustrate this in Fig.~\ref{fig:free_energy_landscape}, where we show the mSST free energy landscape for DG morphology at a fixed set of molecular details, $f=0.3$ and $\epsilon = 1$ and also the grid points in $(p_{1},p_{2})$ where the data was sampled. By varying the terminal boundaries and associated tessellations of DG, we arrive at thermodynamically optimal packing of chains in strong segregation limit. 


\begin{figure}
\center
\includegraphics[width=.8\textwidth]{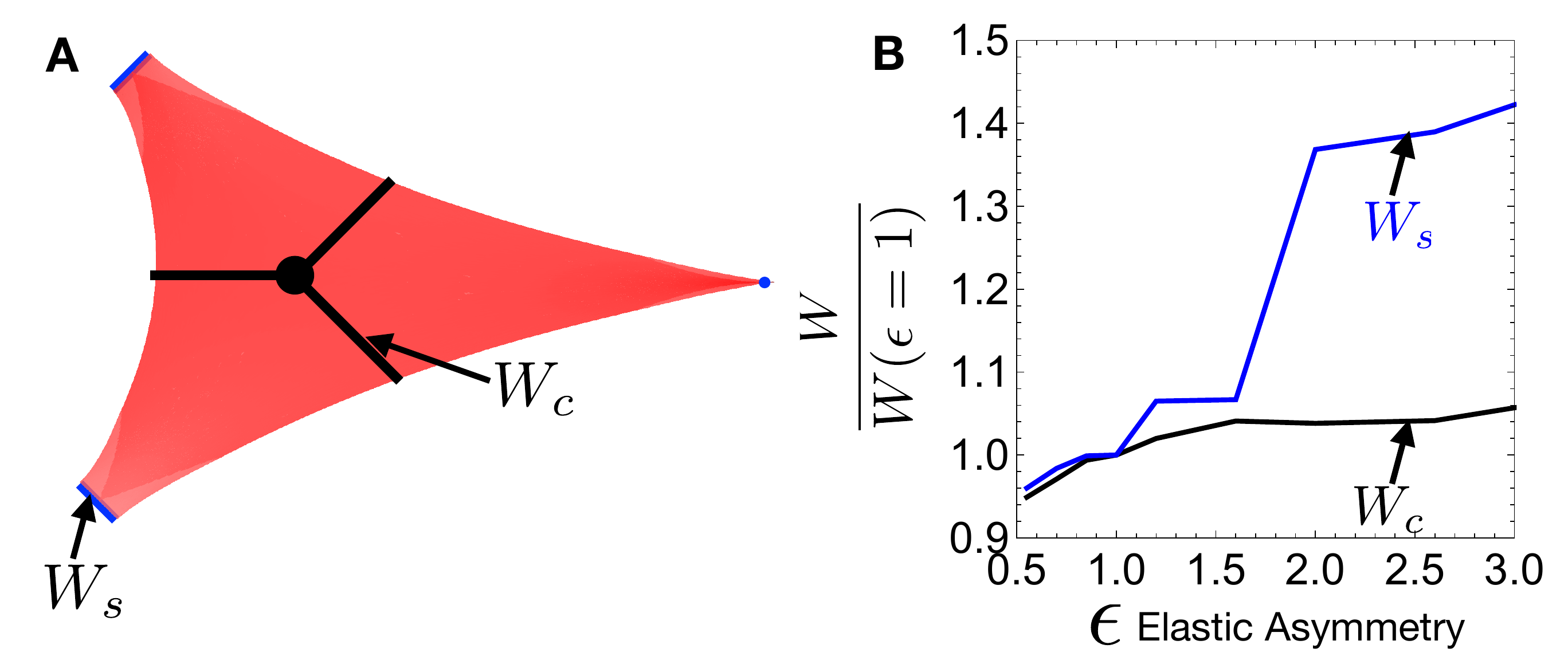}
\caption{\label{fig:web_width} Changing dimensions of the medial web as a function of $\epsilon$. (A) The two metrics of medial web shape we consider are the centroidal width $W_c$ (distance between the centroid and the nearest boundary of the medial web) and the strut width $W_s$ (minimal distance between the two boundary segments along the strut joining neighboring nodes). (B) Both widths, normalized by their $\epsilon = 1$ values, plotted as a function of $\epsilon$. Both dimensions increase with $\epsilon$, but the strut width $W_s$ shows a remarkable transition as $\epsilon \to 2$, indicating a change in the organization of chains in the tubular domain. }
\end{figure}

\section{Skeletal ansatz strong segregation theory}

\label{sec: skeletal}

Here, we outline our approach to constructing a {\it skeletal ansatz} for SST calculations of DG. In particular, we aim to closely recapitulate the prior SST approach of Olmsted-Milner (OM)~\cite{Olmsted1998}, which via a different method, constructed tessellating wedges that span from vertices on the G minimal surface (i.e.~the terminal boundary in the matrix domain) to the 1D skeletal graphs, which is taken as the terminal boundary of the tubular domains.  The OM approach considers a subsequent free energy optimization of the distribution of anchoring points on the skeleton.  

Here, we construct an analog to the OM skeletal structure which is generated from medial tessellations.  In short, the process has three steps: First, generate the medial maps (${\bf T}_{\rm A}$ and ${\bf T}_{\rm B}$) from given generating surface ${\bf G}$.  Second, map the terminal points ${\bf T}_{\rm A}$ onto the skeletal graph that spans the 16b Wyckoff positions of $Ia\bar{3}d$ (it can be shown that these graphs always lie in the medial set of the tubular domains).  Last, balance the wedges and construct the IMDS for area and second-moment calculations.


\begin{figure}
\center
\includegraphics[width=0.75\textwidth]{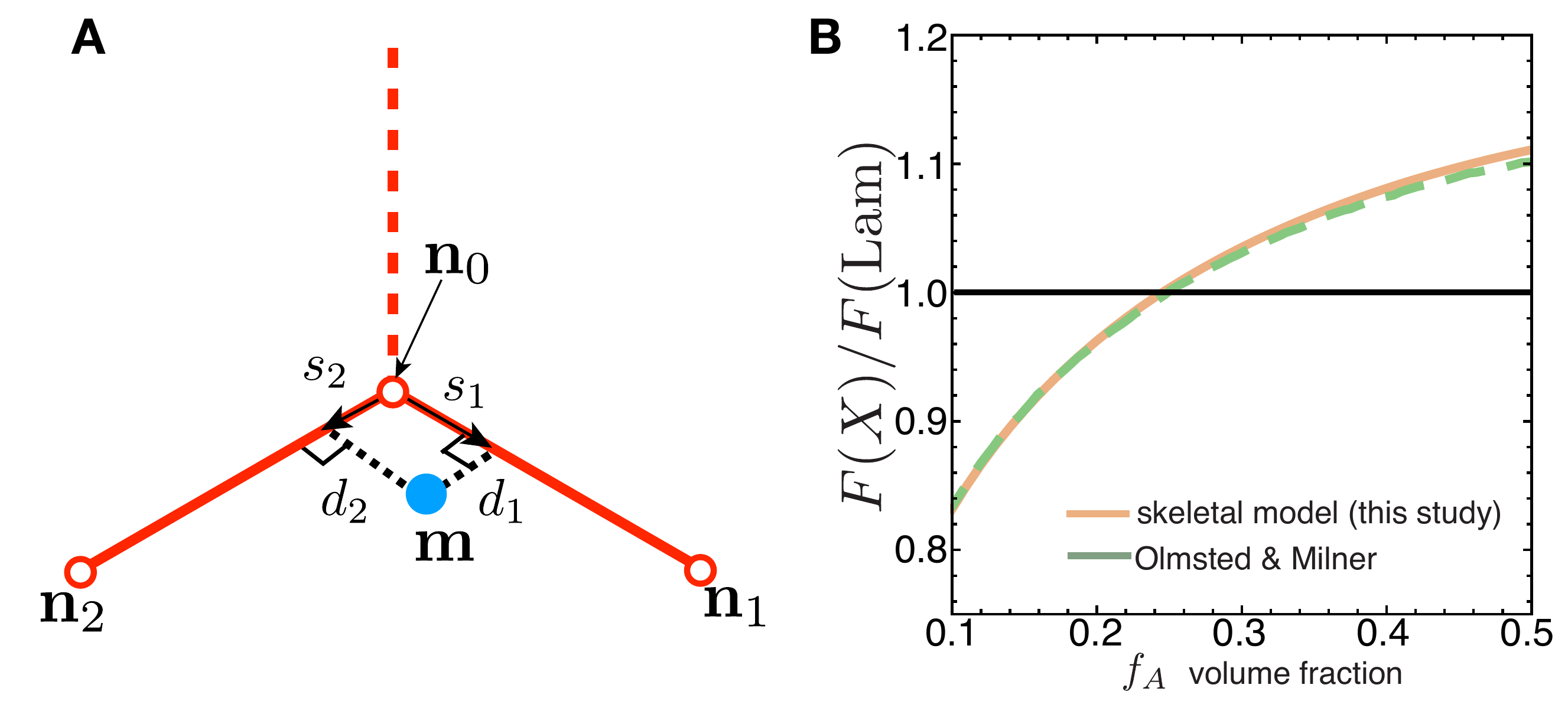}
\caption{\label{fig:smap} (A) Schematic illustrating our procedure to compute skeletal map for each vertex $\mathbf{m}$ on discrete mesh of tubular medial set. (B) Free energy per chain comparison of our skeletal SST calculation for $\epsilon = 1$ with that of Olmsted and Milner \cite{Olmsted1998} }
\end{figure}

The first and last steps of the skeletal SST follow the identical approach used for the medial packings.  Here, we describe the additional algorithm used to map vertices in ${\bf T}_{\rm A}$ to the skeleton ${\bf S}$.  This operation is performed independently for each vertex on ${\bf T}_{\rm A}$, transferring the facet connectivity from the original ${\bf G}$ to the final skeletal map, in effect ``folding'' triangular elements onto the 1D graph.  Our skeletal map proceeds as follows.  Consider a vertex ${\bf m}$ on ${\bf T}_{\rm A}$.  Identify the three closest nodes in the skeletal graph, and denote these as ${\bf n}_0$, ${\bf n}_1$ and ${\bf n}_2$, indexed in order of increasing distance.  This set of nodes correspond to two adjacent edges of the graph ${\bf n}_0 \to {\bf n}_1$ and ${\bf n}_0 \to {\bf n}_2$ (see Fig.~\ref{fig:smap}A).  By definition, ${\bf m}$ is closer to ${\bf n}_0 \to {\bf n}_1$ than ${\bf n}_0 \to {\bf n}_2$.  

Naively, we'd expect an optimal skeletal packing to minimally displace ${\bf m}$ to ${\bf S}$, that is, map onto the closest point on ${\bf S}$.  However, such a map leads to ``jumps'' in the map for ${\bf m}$ close to ${\bf n}_0$ where the closest distance jumps discontinuously from one edge to the next for two nearby points on ${\bf T}_{\rm A}$.  This discontinuity, in turn, leads to gaps and seams in the resulting volume balanced IMDS, resulting in free energies that far exceed the upper bounds of the OM skeletal ansatz.  

To achieve a map from ${\bf T}_{\rm A}$ to the skeleton ${\bf S}$ that is smooth in the neighborhood of the junction points (corresponding to wedges located in the ``elbow'' region of DG), we use a map that interpolates between the two closest locations on the two closest edges for a given point.  In practice we do this by parameterizing the spans between ${\bf n}_0 \to {\bf n}_m$ (where $m=1,2$) as
\begin{equation}
    {\bf e}_m (s_m) = {\bf n}_0 + s_m ({\bf n}_m-{\bf n}_0) ,
\end{equation}
where $s_m\in[0,1]$ is a parameter that spans the edge. The point of closest contact $s^*_m$ along the $m$ edge is
\begin{equation}
 s^*_m = \left\{ \begin{array}{ll} 0 & {\rm for \ } ({\bf m} - {\bf n}_0) \cdot ({\bf n}_m - {\bf n}_0)<0 \\ 
 \frac{({\bf m} - {\bf n}_0) \cdot ({\bf n}_m - {\bf n}_0)}{|{\bf n}_m - {\bf n}_0|^2} & {\rm for \ } 0\leq ({\bf m} - {\bf n}_0) \cdot ({\bf n}_m - {\bf n}_0) \leq |{\bf n}_m - {\bf n}_0|^2 \\ 
 1 & ({\bf m} - {\bf n}_0) \cdot ({\bf n}_m - {\bf n}_0) \leq |{\bf n}_m - {\bf n}_0|^2> 1 {\rm for \ }
 \end{array} \right. ,
\end{equation}
and the corresponding closest distance to those edges is $d_m = |{\bf m} - {\bf e}_m(s_m^*)|$.  Based on these distances, our map associates ${\bf m}$ to the region spanned by the two nearest edges by constructing a composite coordinate $\sigma$ such that $S = s_1$ for $S\geq 0$ (and maps along ${e}_1$) and $S=-s_2$ for $S< 0$ (and maps along ${\bf e}_2$).  In this way, we determine the mapped location of ${\bf m}$ along this span $\sigma({\bf m})$ by a weighted average of the closest points
\begin{equation}
    \sigma({\bf m}) = \frac{s_1 w(d_1)-s_2 w(d_2)}{w(d_1)+w(d_2)} ,
\end{equation}
where $w(d) = d^{-q}$ are weighting functions of distance.  It is straightforward to show that $S({\bf m}) \geq 0$ by construction, and the skeletal map is always along the closest edge (unless at the node ${\bf n}_0$ itself), so that the map to ${\bf S}$ is given by
\begin{equation}
    {\bf S}({\bf m}) = {\bf e}_1 \big( \sigma({\bf m}) \big) .
\end{equation}
To further optimise the distribution of mapped points onto ${\bf S}$ we consider a range of the exponents $q = 1,2,3$, and select the value (along with the optimal generating surface) that minimizes the SST free energy of skeletal DG for a given set of $(f, \epsilon)$. In our study, we find that $q=2$ solutions have lowest energy compared to the other exponents and we compare the free energy per chain in Fig.~\ref{fig:smap}B with that of Olmsted \& Milner's skeletal model, which we read off Fig.~16 of ref. \cite{Olmsted1998}. At $\epsilon=1, f=0.3$, $F(\rm{DG})$ from the skeletal model studied in this work is 0.4\% higher than Olmsted \& Milner's result. In comparison with $F(\rm{DG})$ at the same $f$ from mSST, the relative difference in free energy between medial and skeletal SST is 3.1\%. 

\section{SST model of Hex phase}

\label{sec: hex}

To compute the SST free energy of the competitor Hex phase we follow the method of ref.~\cite{Grason2004}.  Here, we consider a model where the columns are packed in a hexagonal lattice and the terminal boundaries of the outer matrix phase are composed of the hexagonal Voronoi cells, while we assume that the the inner terminal boundary of the core block is a 1D line in the center of the column.  It is well established that the lattice packing of the columnar phase introduces at least some measure of packing frustration due to fact that rays extending from the center of the column extend variable distances to reach the hexagonal outer terminal boundary.  Depending on the relative distance of the IMDS from the outer terminal boundary as well as the ratio of the coronal to core stiffness (parameterized by $\epsilon$), this can lead to warping of the IMDS shape away from the constant curvature cylindrical shape.  

To account for this, the SST ansatz of ref.~\cite{Grason2004} considers a variable class of IMDS shapes, controlled by a variational parameter $\alpha\in [0,1]$, that interpolates continuously between a circular cross section ($\alpha =0$) to a perfect hexagonal cross section ($\alpha =1$) at fixed enclosed core area fraction.  While the inner (A) blocks extend radially from the cell center, volume balance and the variable coronal thickness requires the outer blocks to tilt relative to the radial direction, by some degree that varies with position and interface shape (degree of tilting increases for round IMDS shapes that approach the outer terminal boundary and vanishes for IMDS shapes that are affine copies of the hexagonal cell shape).

\begin{figure}
\center
\includegraphics[width=.8\textwidth]{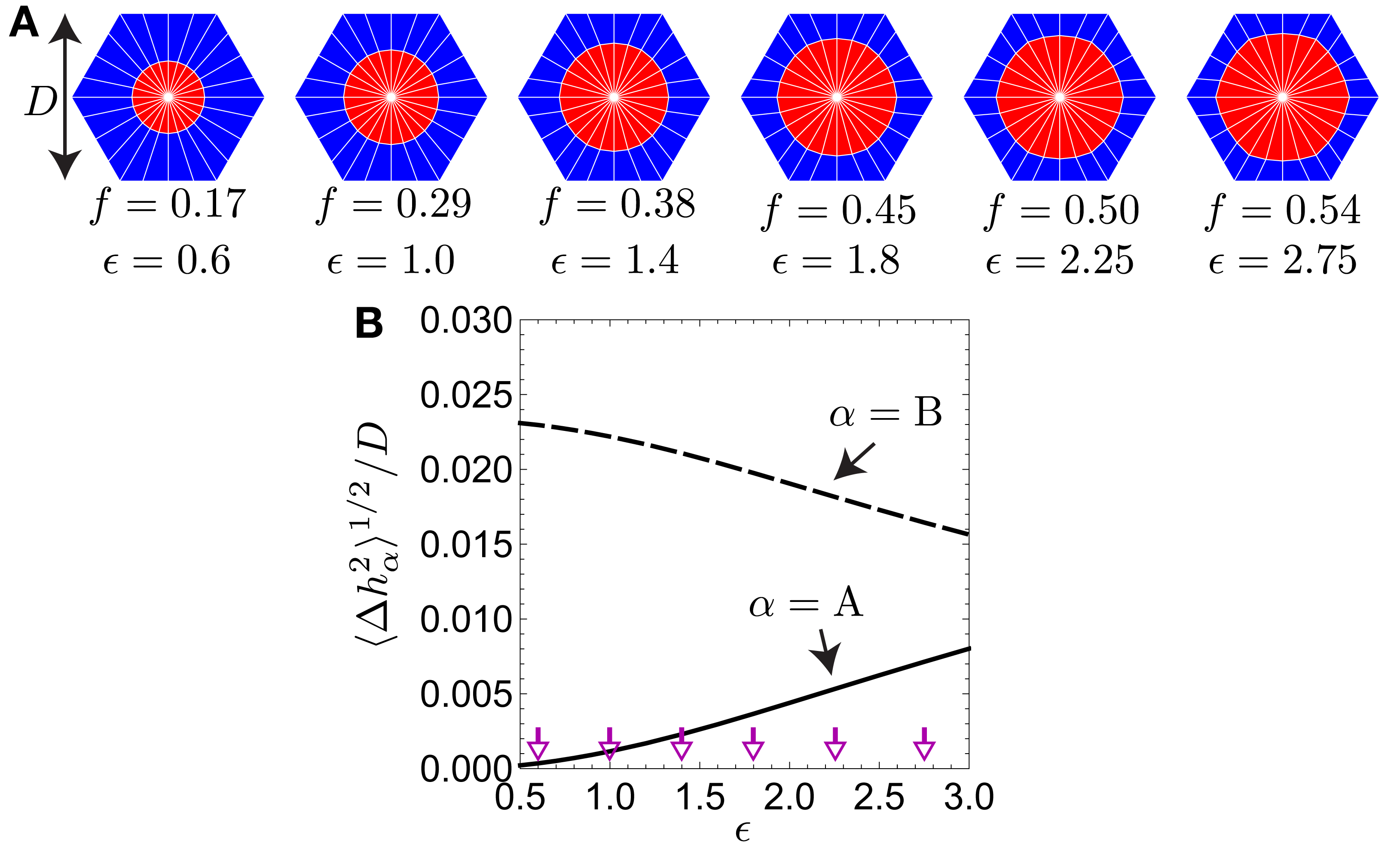}
\caption{\label{fig:hex} (A) Illustrations of a single Voronoi cell of the Hex phase with IMDS distorted by packing frustration. The number of wedges used to tessellate each cell has been reduced for clarity. The $\alpha$ is the distortion parameter that minimizes the free energy at fixed $f$. (B) Variance in the thickness $h_\alpha$ of block A (solid) and block B (dashed) as a function elastic asymmetry parameter $\epsilon$.}
\end{figure}

The variational calculation for Hex minimizes the SST free energy over the IMDS shape, providing an upper bound (within the PBT approximation of brush entropy) of the equilibrium free energy.  In contrast to the tessellation of DG into a finite (but very large) number of wedge volumes, the free energy in this ``kinked path'' ansatz for Hex can be evaluated without explicit discretization of the geometry, by numerical integration executed in Mathematica.  Fig.~\ref{fig:hex}A shows the variation of optimal Hex packings along the boundary where $F({\rm Lam})=F({\rm Hex})$, showing the transformation from circularly IMDS shapes for $\epsilon\lesssim 2$ (and low $f$) to more polygonal IMDS shape for $\epsilon\gtrsim 2$ (and higher $f$).  This subdomain morphology transition is driven by increasing effects of packing frustration in the outer coronal strongly, favoring a more uniform packing at the expense of increased cost of non-circular IMDS shape and core block stretching.  In Fig.~\ref{fig:hex}B, we show the root-mean-square variance of A and B block extension for these equilibrium morphologies along the $F({\rm Lam})=F({\rm Hex})$ line as function of elastic asymmetry $\epsilon$.

\section{Surface shape analysis}

We have developed a set of computational tools to analyze the shape of each IMDS.
These tools rely on the construction of the IMDS as a triangulated mesh: an ordered collection of points $\{\mathbf{v}_i\}_{i=1}^{N_v}$, where $N_v$ is the number of vertices, that is grouped into collections of triangular facets $\{f_\mu\}_{\mu=1}^{N_f}$, where $N_f$ is the number of facets.

\subsection{IMDS curvature calculation}

Smooth surfaces embedded in 3 dimensions are characterized by two types of curvature: Gaussian curvature $K_G$ and mean curvature $H$.
When these surfaces are approximated by meshes, there is no unique definition of the discrete analogues of these curvatures.
Instead, there are ways of defining discrete analogues of curvature such that the smooth definitions of curvature are attained in the limit of an infinitely fine mesh \cite{Crane2017}.
While the generating surfaces are formed from well-refined meshes, after the process of volume balance and joining the wedge dividing planes together, the resulting IMDS does not necessarily have a well-behaved mesh for discrete definitions of curvature; this is especially true for those generated from the skeletal ansatz. 
Thus, we adopt method of calculating curvatures based fitting smooth surfaces to point clouds, a technique that has proven to produce smooth curvature distributions on complex, potentially poorly meshed surfaces \cite{Krishnamurthy1996,Douros2002}.

To start, select a vertex $\mathbf{v}_i$ and a collection of $\mathcal{N}$ nearest points to that vertex, as determined by the Euclidean distance.
This neighborhood of $\mathbf{v}_i$ defines a surface patch $\mathcal{P}_i$ of the meshed surface.
From the selection of points $\mathbf{v}_m \in \mathcal{P}_i$, where $m = 1,\dots,\mathcal{N}$, we determine the center of mass $\mathbf{v}_i^* = \mathcal{N}^{-1}\sum_{m}\mathbf{v}_m$ and inertia tensor $\mathcal{I}_i = \mathcal{N}^{-2}\sum_{m,n}(\mathbf{v}_m-\mathbf{v}_i^*)\otimes(\mathbf{v}_n-\mathbf{v}_i^*)$.
We use the eigenvectors of this inertia tensor to construct a local orthonormal basis $\{\mathbf{\hat{e}_1},\mathbf{\hat{e}_2},\mathbf{\hat{e}_3}\}$, where we select $\mathbf{\hat{e}_3}$ to align with the minimum eigenvalue of $\mathcal{I}_i$, representing the direction of minimal vertex variation.
We represent each point $\mathbf{v}_m \in \mathcal{P}_i$ in the patch via a center of mass coordinate system, $\mathbf{v}_m - \mathbf{v}_i^* = u_m \mathbf{\hat{e}_1} + v_m \mathbf{\hat{e}_2} + w_m \mathbf{\hat{e}_3}$.

Next, given the collection of points $(u_m,v_m,w_m)$, we fit a quadric form
\begin{equation}
    h(u,v) = a_1 u^2 + a_2 v^2 + a_3 u v + b_1 u + b_2 v + c
\end{equation}
via least squares by minimizing the sum $ \sum_{m}\left|h(u_m,v_m) - w_m\right|^2$ over the free parameters $a_1$, $a_2$, $a_3$, $b_1$, $b_2$, and $c$.
The result is a smooth approximation $\mathbf{r}_i(u,v) = u \mathbf{\hat{e}_1} + v \mathbf{\hat{e}_2} + h(u,v) \mathbf{\hat{e}_3}$ of the surface patch $\mathcal{P}_i$ in the Monge representation \cite{Kamien2002}.
Using the Monge representation, the Gaussian curvature is given by
\begin{equation}
    K_G(u,v) = \frac{(\partial_{uu}h) (\partial_{vv}h) - (\partial_{uv}h)^2}{\left[1 + (\partial_u h)^2 + (\partial_v h)^2\right]^2}
\end{equation}
and the mean curvature is given by
\begin{equation}
    H(u,v) = \frac{[1 + (\partial_v h)^2]\partial_{uu}h + [1 + (\partial_u h)^2]\partial_{vv}h - 2 (\partial_u h)(\partial_v h)\partial_{uv}h}{2\left[1 + (\partial_u h)^2 + (\partial_v h)^2\right]^{3/2}} \, .
\end{equation}
Evaluations of $K_G(u,v)$ and $H(u,v)$ yield approximate curvatures for each point $\mathbf{v}_m \in \mathcal{P}_i$: $K_{G,m} = K_G(u_m,v_m)$ and $H_m= H(u_m,v_m)$.

\begin{figure}
\center
\includegraphics[width=.8\textwidth]{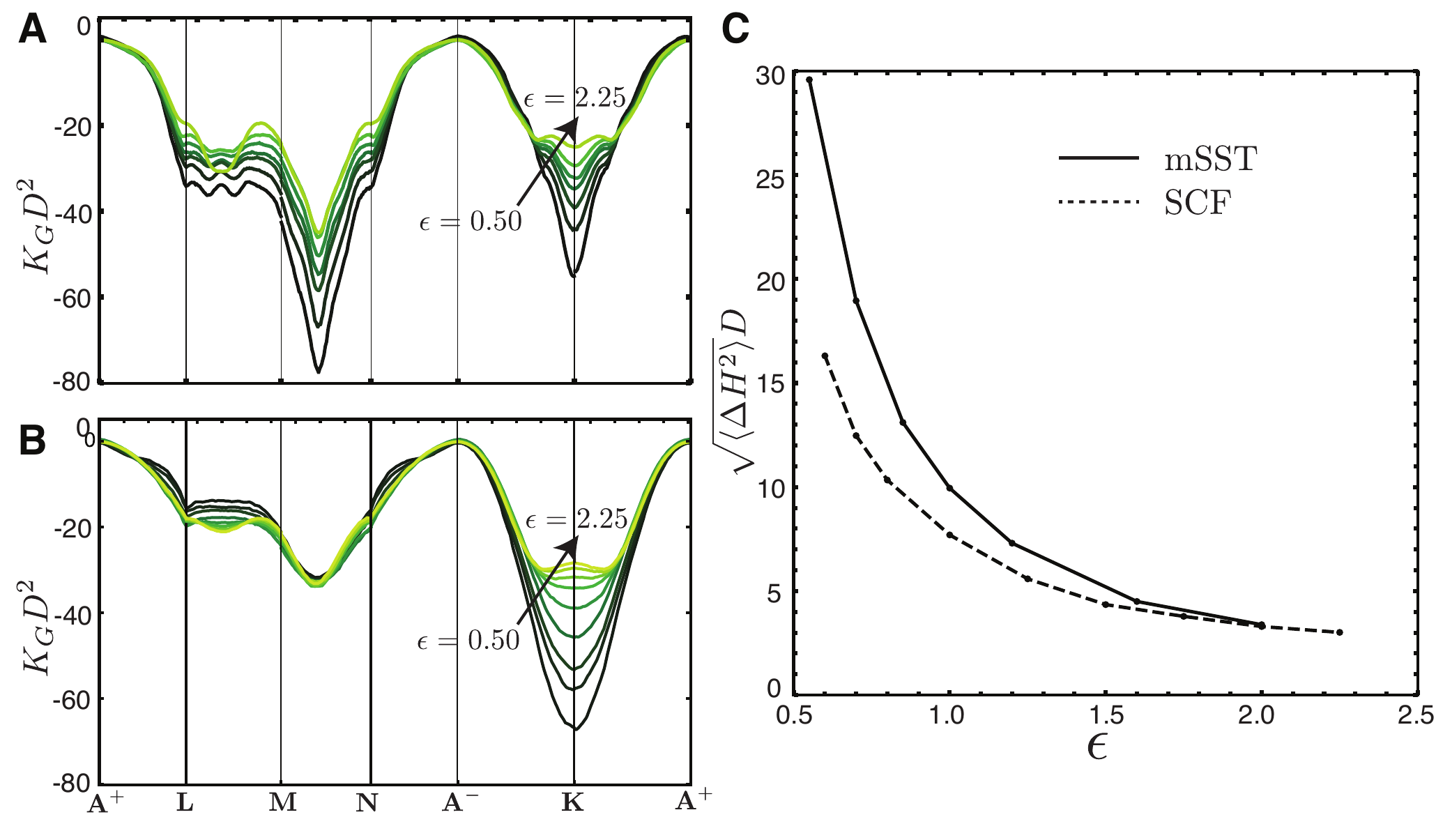}
\caption{\label{fig:combined_curvature} Supplement to Fig.~5. Band diagrams showing the variation of Gaussian curvature along the path defined in Fig.~1D are shown for IMDSs generated by (A) mSST and (B) SCF. (C) shows the variance in mean curvature as a function of elastic asymmetry $\epsilon$.}
\end{figure}

We repeat this process for every point $\mathbf{v}_i$ in the mesh.
The collection of surface patches $\bigcup_i \mathcal{P}_i$ yields an atlas that covers the meshed surface.
Since every surface patch contains $\mathcal{N}+1$ points, each of which generates its own surface patch, each surface patch intersects at least $\mathcal{N}$ other surface patches, yielding many measurements of the surface curvature.
Define $K_{G,j|i}$ to be the Gaussian curvature of point $\mathbf{v}_j$ as measured in patch $\mathcal{P}_i$; there is a similar definition for mean curvature.
We then define an accepted value $\overline{K_G}_{,j}$ as the \emph{average} of $K_{G,j|i}$ over all patches $\mathcal{P}_i$ containing $\mathbf{v}_j$; we similarly define the accepted mean curvature $\overline{H}_j$.
While the choice of neighborhood size $\mathcal{N}$ yields different resolutions of curvature distribution, we find that for our meshes, $\mathcal{N}=50$ yields a reasonable trade-off between measuring a smooth curvature distribution that ``averages-out'' features due to the discreteness of the mesh and nonetheless retaining fine features of the curvature distributions.

\subsubsection{Skeletal IMDS curvature}

For the IMDS generated from skeletal SST, we introduce additional steps to smooth the mesh before measuring a curvature distribution.
We find this to be necessary since the association map between mesh elements and the gyroid skeleton is not smooth in the vicinity of a node.
Proceeding with the patch fitting algorithm discussed above, the mesh undergoes two pre-processing steps.
We start by approximating the normal vector at each vertex in the mesh as the average of the normal vectors of the surrounding facets.
The angle between the normal vectors of neighboring faces gives an estimate of mesh curvature at the scale of a facet.
We identify the roughest regions of the mesh by selecting vertices that have a maximum nearest neighbor normal vector change greater than $\pi/8$.
The coordinates of these problem vertices are then averaged with their nearest neighbors.
We then identify the longest edges in the mesh, thresholding by the average edge length plus the standard deviation.
From this list of edges, we identify vertices with at least 3 long edges and average the coordinates of those vertices with any neighboring vertices that are not in the list of ``problem'' vertices.

\begin{figure}
\center
\includegraphics[width=.6\textwidth]{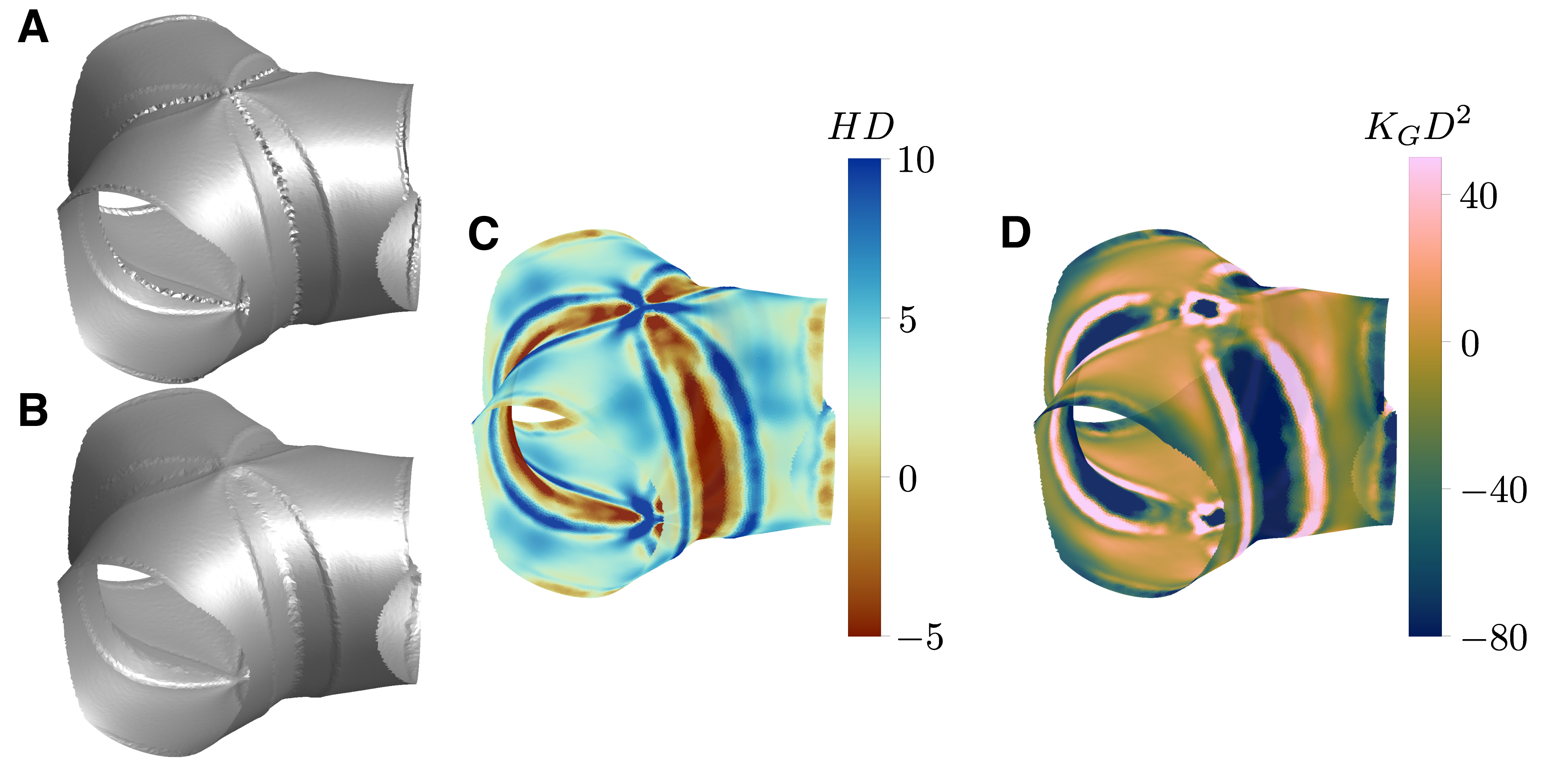}
\caption{\label{fig:skeletal_smoothing} Mesh processing for skeletal IMDS, (A) pre-smoothing and (B) post-smoothing. (C) Mean curvature and (D) Gaussian curvature distributions of the smoothed mesh.}
\end{figure}

After these pre-processing steps, we use the same patch fitting algorithm for calculating curvature to further smooth the mesh.
Since each surface patch has an associated quadratic fit, $h(u,v)$, we can use the quadratic fit for a given patch $\mathcal{P}_i$ to adjust each vertex $\mathbf{v}_j$ to a new position $\mathbf{\tilde{v}}_j$ that lies on the smooth approximation of the surface patch.
A given coordinate $\mathbf{v}_j$ exists in multiple surface patches and thus has multiple smooth images; we shall denote the image of $\mathbf{v}_j$ under the map induced by the quadratic fit to patch $\mathcal{P}_i$ as $\mathbf{\tilde{v}}_{j|i}$.
Much like the averaging process for determining surface curvatures, we map each vertex $\mathbf{v}_j$ to the average of $\mathbf{\tilde{v}}_{j|i}$ over all containing patches.

We can iterate the pre-processing and mesh smoothing steps as many times as necessary.
Figure \ref{fig:skeletal_smoothing} shows the result after four iterations of the patch fitting algorithm.

\subsection{``Band'' diagrams}

In our analysis, we introduce a number of quantities, such as such as domain thickness, curvature, and chain tilt, that vary over a given IMDS.
We have found it useful to plot these quantities along a closed path on the IMDS that passes through various symmetry points, much like electronic band structure diagrams.
Here, we outline this plotting method.

\begin{figure}
\center
\includegraphics[width=.4\textwidth]{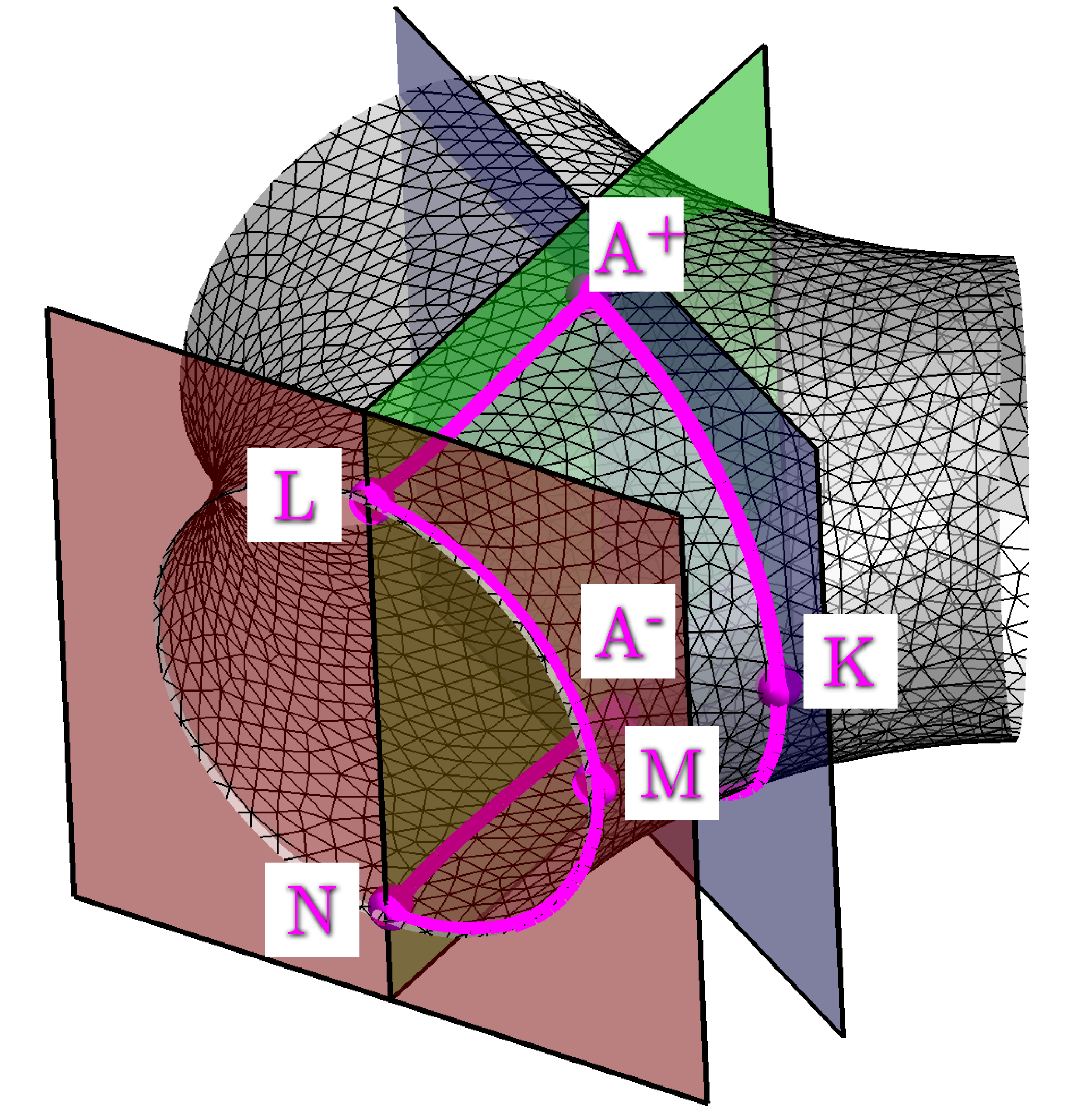}
\caption{\label{fig:node_cutting} The three slicing planes passing through symmetry points of the DG nodal region used to construct the ``band'' diagrams used to plot various quantities that vary over the IMDS.}
\end{figure}

Let $f(\mathbf{r})$ be a scalar field that takes values at points $\mathbf{r}$ on an IMDS and let $\bm{\gamma}(s)$ represent a closed path on the surface of the IMDS that passes through symmetry points to be defined below.
Here, $s$ is a parameter that varies from $0$ to $s_{\rm max}$ and $\bm{\gamma}(L) = \bm{\gamma}(0)$ for a closed loop.
The resulting ``band'' diagram is a plot of $f$ along $\bm{\gamma}$, namely $f(s) = f\circ\bm{\gamma}(s)$.
While we can, in general, choose any path along the surface, we choose a path that passes through a selection of symmetry points.
Each segment of the patch $\gamma$ can be represented as the intersection of the IMDS with a plane that joins sets of symmetry points:
\begin{itemize}
    \item \textbf{A}\textsuperscript{+} and \textbf{A}\textsuperscript{-} lie along [111] and pierce the nodal region, forming an axis of 3-fold rotational symmetry.
    \item \textbf{K} is located at a point of 2-fold rotational symmetry, on the saddle-like region between two struts.
    \item \textbf{L}, \textbf{M}, and \textbf{N} lie on the plane that bisects a strut. \textbf{L} and \textbf{N} are co-planar with \textbf{A}\textsuperscript{+} and \textbf{A}\textsuperscript{-}.
    \item \textbf{M} and \textbf{K} lie on a plane that cuts transversely through the 3-fold symmetry axis.
\end{itemize}
Therefore, the path \textbf{A}\textsuperscript{+} $\to$ \textbf{L} $\to$ \textbf{M} $\to$ \textbf{N} $\to$ \textbf{A}\textsuperscript{-} $\to$ \textbf{K} $\to$ \textbf{A}\textsuperscript{+} can be obtained from the intersection of the IMDS with three cutting planes shown in Fig.~\ref{fig:node_cutting}: (i) the plane formed by the points \textbf{A}\textsuperscript{+}, \textbf{L}, \textbf{N}, \textbf{A}\textsuperscript{-}, (ii) the plane formed by the points \textbf{L}, \textbf{M}, \textbf{N}, and (iii) the plane formed by the points \textbf{A}\textsuperscript{+}, \textbf{K}, \textbf{A}\textsuperscript{-}.

To create continuous paths from the intersection of each cutting plane with the mesh, we first identify every mesh edge that pierces through a cutting plane.
These edges are ordered sets of vertices, $e_{ij} = (\mathbf{v}_i,\mathbf{v}_j)$, which each form a line $\mathbf{e}_{ij}(\lambda) = (1-\lambda)\mathbf{v}_i + \lambda\mathbf{v}_j$, for $\lambda \in [0,1]$.
For each cut edge $e_{ij}$, we determine the value $\lambda^*$ that intersects with the plane, using the formula
\begin{equation}
    \lambda_{ij}^* = \frac{-(\mathbf{v}_i - \mathbf{p}_{\rm plane})\cdot\hat{\mathbf{n}}_{\rm plane}}{(\mathbf{v}_j - \mathbf{v}_i)\cdot\hat{\mathbf{n}}_{\rm plane}} \, ,
\end{equation}
where $\mathbf{p}_{\rm plane}$ is a point that lies in the cutting plane and $\hat{\mathbf{n}}_{\rm plane}$ is the plane's unit normal vector.
The list of plane-mesh intersection points $\mathbf{e}_{ij}(\lambda^*_{ij})$ is then ordered by the Euclidean distance of each point from a given end point of the path (for example, we choose $\bm{\gamma}(0)$ to lie at \textbf{A}\textsuperscript{+}).
This process results in an ordered collection of points $\bm{\gamma}_i$ that are linear interpolations of mesh vertices.

Most of the data $f(\mathbf{r})$ that we plot are defined on mesh facets.
To evaluate on the path, we first find approximate values of $f$ at mesh vertices by averaging over the evaluations of $f$ at facets surrounding a given vertex.
We then use linear interpolation to find values of $f_i = f(\bm{\gamma}_i)$ that lie along cut edges.
The result is an ordered collection of values $\left(\bm{\gamma}_i,f_i\right)$.
Finally, we make this path continuous over the entire mesh via linear interpolation.

To standardize the length between each path segment, we choose a ``standard'' ruler for path lengths.
Our standard ruler is the level set $\sin(2\pi x)\cos(2\pi y) + \sin(2\pi y)\cos(2\pi z) + \sin(2\pi z)\cos(2\pi x) = 1$, which has interval lengths $s_{\rm\bf A^+ L} = s_{\rm\bf NA^-} = 0.177$, $s_{\rm\bf LM} = 0.193$, $s_{\rm\bf MN} = 0.183$, and $s_{\rm\bf A^- K} = s_{\rm\bf K A^+} = 0.237$.

\section{Chain tilt at the IMDS}
\label{sec: tilt}
The medial map as defined in eq.~\ref{medialmap} ensures that the normal $\mathbf{\hat{N}_{\rm G}}$ to a triangular facet belonging to $\mathbf{G}$ is parallel to the centroidal vector $\mathbf{C}$. 
However, the normal $\mathbf{\hat{N}_{\rm I}}$ of the dividing plane at a given wedge is not generally collinear with $\mathbf{C}$ but makes an angle $\theta$ where $\mathbf{\hat{N}_{I}}\cdot\hat{\mathbf{C}} = \cos\theta$.
Within mSST, we assume that the chain extension is along straight lines connecting the centroids of the faces on the IMDS to those on the terminal boundary $\mathbf{T}_{\alpha}$.

To test the validity of medial packing we compare the IMDS tilt pattern from mSST to SCFT calculations at finite $\chi N (=75)$.  We compute SCFT solutions for DG from the PSCF code \cite{Arora2016} and use these solutions to compute both the shape of the IMDS as well as the vector order parameter ${\bf p}({\bf x})$ that points along the mean field trajectory of chains using the methods described in \cite{Prasad2017}.  Briefly, this order parameter derives from derivatives of the segment distribution functions $q(n,{\bf x})$ and $q^\dag (n,{\bf x})$ which describe the statistical weight of chain segments extending from their free ends at their A and B blocks, respectively, to the $n^{\rm th}$ segment at point ${\bf x}$.  Given mean-field solutions for these distribution functions, the segment polar order parameter for the $\alpha^{\rm th}$ component is given by 
\begin{equation}
    {\bf p}_\alpha ({\bf x}) = -\frac{V a_\alpha }{N} \int_{n\in \alpha} {\rm d}n ~ \Big[ q^\dag \nabla q - q \nabla q^\dag \Big],
\end{equation}
where we have directed the orientation from the B to the A ends, to match to convention of the mSST trajectories.  We extract the end distributions from converged solutions of PSCF and compute these derivatives numerically, arriving at the polar order parameter profile of both blocks, from which we compute the segment averaged orientation ${\bf p} ({\bf x}) = {\bf p}_{\rm A} ({\bf x})  +{\bf p}_{\rm B} ({\bf x})$ which can be converted to a unit vector to map the mean direction of chain trajectories, ${\bf \hat{p}} ({\bf x})  $.  As in the previous section, we extract the IMDS from the level set condition $\phi_{\rm A} ({\bf x}) = 1/2$ and its local normal at these points by ${\bf N} ({\bf x}) = \nabla \phi_{\rm A} /|\nabla \phi_{\rm A}|$.  From these two fields, we compute and map the local tilt profile from $\cos \theta (\mathbf{x}) = {\bf \hat{p}} ({\bf x}) \cdot {\bf N} ({\bf x})$ at the IMDS.

\begin{figure}
\center
\includegraphics[width=.8\textwidth]{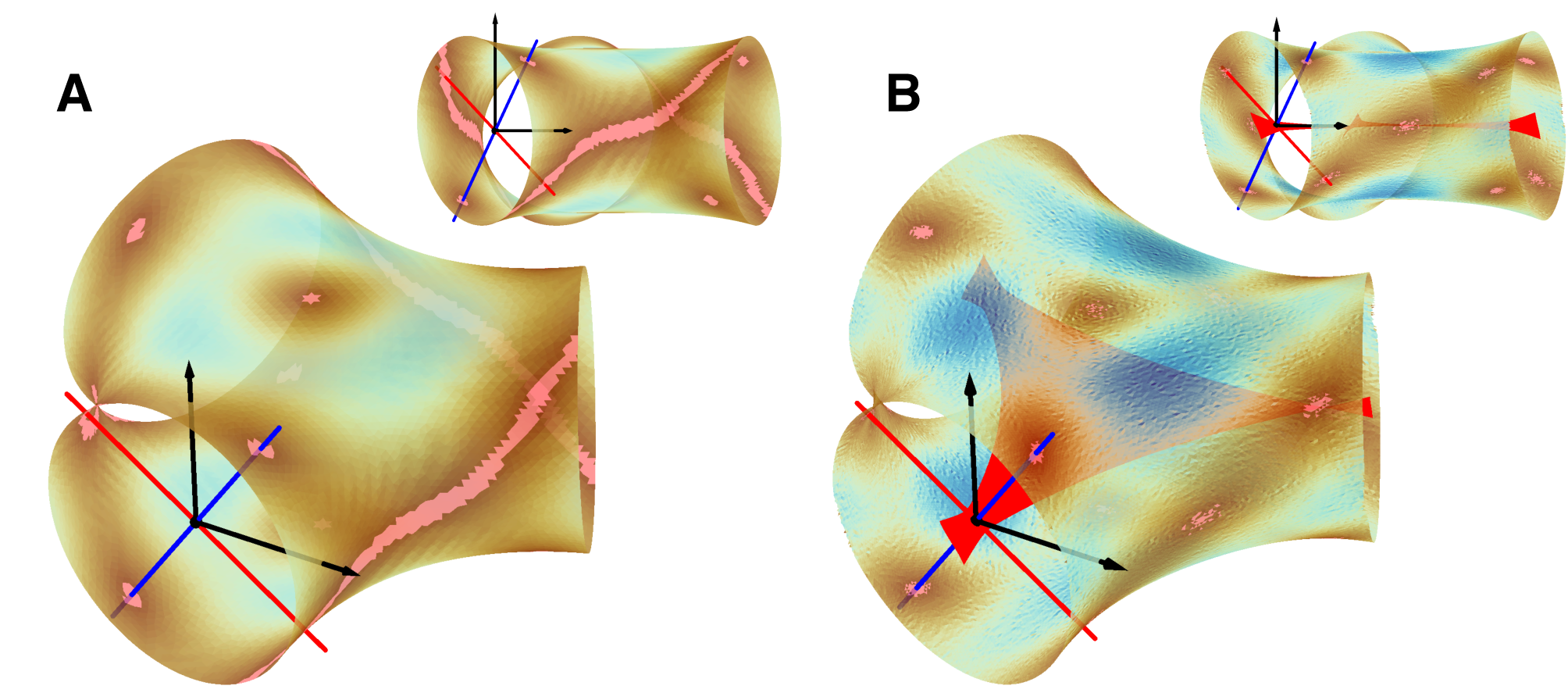}
\caption{\label{fig:tilt} Map of chain tilt $\theta$ on the IMDS from (A) SCF and (B) mSST. Regions of near-normal chain orientations $0 \leq \theta \leq 1^\circ$ are highlighted in pink. Red and blue line segments show the $70.5^\circ/2$ twist of the IMDS between neighboring nodal regions, oriented tangent and normal to the medial set, respectively. The passage of the medial tangent and normal through locations of near-normal chain orientations is clarified in the rotated view shown in insets.}
\end{figure}

In Fig.~\ref{fig:tilt} we color the volume balanced IMDS with the local tilt $\theta$ within each wedge, here shown for case of $f=0.3$ and $\epsilon=1$ for the free energy mSST packing.  We also compare this to result of finite $\chi N = 75$ from SCFT in the same figure.
The patterns of non-zero tilt implies that the medial sets of \textbf{G} are not the medial sets of the IMDS, and hence chain trajectories are not strictly medial lines.
However, there are regions where this tilt in minimal and approaches $0^\circ$, where the chain extension is locally medial.
These near normal regions are highlighted in pink in Fig.~\ref{fig:tilt} and are remarkably similar in pattern between SCF and mSST.
In particular, half-way along the strut joining the highlighted nodal region to a neighboring region, we find that these low-tilt regions are rotated relative to the nodal axis (the [111] direction), points ${\bf A}^{\pm}$, which is consistent with the fact that 3-fold symmetry through this axis.  Likewise, the tilt vanishes at point ${\bf K}$ which is the location of 2-fold axis, whose symmetry also requires vanishing of tilt at the IMDS.  Points of medial or near-medial packing are also distributed in a non-trivial pattern away from these symmetry points.  Most notably, a satellite spot highlighted midway along the strut whose angular location (rotated by $70.5^\circ/2$ relative to the plane of the strut) indicates the coincidence of the IMDS normal and the normal to the underlying medial web.  In this way, we may view this satellite spot of vanishing tilt as an image of the local orientation of the underlying terminal surface, a feature which is also reflected in the tilt pattern of the SCFT solution.
Similarly we observe, a low-tilt ``scar'' shown in both structures, intersects with the strut-bisecting plane (the {\bf LMN}-plane, as defined in our band diagrams) and corresponds to trajectories that terminate at the {\it edge} of the terminal web.  We take the high correspondence between the tilt-patterns predicted by mSST and SCFT (particularly at off-symmetry locations) as a strong evidence that our medial construction accurately captures the underlying and non-trivial patterns of chain packing in finite segregation DG morphologies.

\section{End distributions comparisons}

\label{sec: end}

Here, we describe the comparison of end distributions for DG structures between medial and skeletal SST calculations and SCFT at finite $\chi N =75$.  In particular, we focus on the distribution of A block ends $\rho_{\rm A} ({\bf x})$ and the extend to which it concentrates on or spreads away from the skeletal graphs in the tubular domains.  For SCFT the end distribution is directly proportional to chain end distribution, propagating from the free B end along the full length of the chain to the A end at $n=0$, i.e.  $\rho_{\rm A} ({\bf x}) \propto q^\dag (n=0,{\bf x})$, a quantity which can be directly extracted from PSCF solutions.

The free end distributions for SST come from parabolic brush theory (PBT) \cite{Milner1988}.
PBT only strictly holds for converging (or concave) brushes, i.e.~the brushes where chain ends converge towards a central point in space, which is sufficient for our analysis of $\rho_{\rm A} ({\bf x})$ in the tubular domains.
Converging brushes are defined by an area distribution $A(\tilde{z}) = A_0\mathcal{A}(\tilde{z})$ that decreases monotonically as a function of the height $\tilde{z}$ above the grafting interface of the brush.
Here, $A_0$ is the area element of the grafting interface (the IMDS) and $\mathcal{A}(\tilde{z})$ is a dimensionless function that has a quadratic form $\mathcal{A}(\tilde{z}) \equiv 1 + 2\tilde{H}\tilde{z} + \tilde{K}\tilde{z}^2$, per Steiner's law \cite{Hyde1997_ch4}.
For smooth surfaces, $\tilde{H}$ is identified as the mean curvature $H$ and $\tilde{K}$ is identified as the Gaussian curvature; for the wedges involved in the medial construction presented here, $\tilde{H}$ and $\tilde{K}$ are the linear and quadratic parts of the area function.
For a brush of total height $h$, PBT predicts a free end probability distribution \cite{Semenov1985,Milner1988} given by
\begin{equation}
    g(\tilde{z}) = \frac{1}{h \mathcal{V}(h)}\left\{\mathcal{A}(h)\frac{\tilde{z}/h_{\rm A}}{\sqrt{1 - (\tilde{z}/h)^2}} + 2\tilde{H} \tilde{z} \log\frac{\tilde{z}/h}{1 + \sqrt{1-(\tilde{z}/h)^2}} - 2 \tilde{K} h \tilde{z} \sqrt{1 - (\tilde{z}/h)^2}\right\} \, ,
    \label{eq: gz}
\end{equation}
where $\mathcal{V}(h) \equiv 1 + \tilde{H}h + \tilde{K}h^2/3$ is a dimensionless measure of curved brush volume.

In order to quantify the end distributions of A block chains, we set $\tilde{z} = z - h_{\rm d}$ as the coordinate measuring height with respect to the IMDS.
The area function of eq.~(\ref{eq: Az}) is then re-cast in this shifted coordinate, prompting new values for $A_0$, $\tilde{H}$, and $\tilde{K}$:
\begin{equation}
    \begin{split}
        A_{\mu,0} &\mapsto A'_{\mu,0} = A_{\mu,0}(1 + 2\tilde{H}h_{\rm d} + \tilde{K}h_{\rm d}^2) \\
        \tilde{H} &\mapsto \tilde{H}' = \frac{\tilde{H} + \tilde{K}h_{\rm d}}{\mathcal{A}(h_{\rm d})} \\
        \tilde{K} &\mapsto \tilde{K}' = \frac{\tilde{K}}{\mathcal{A}(h_{\rm d})}
    \end{split}
\end{equation}
Note that the linear end distribution in the form of eq.~(\ref{eq: gz}) is normalized such that $\int_0^h {\rm d}z\, g(z) = 1$, such that it is related to the spatial end distribution by additional factors relating to the number of chains per wedge and local area.


Using the fact that the number of chains within wedge $\mu$ is $\rho V_\mu(h)/N$ and the local area element at $z$ is $A_\mu(z)$ we have 
\begin{equation}
    \rho_{\rm A} (\tilde{z}) = \frac{\rho V_\mu(h)}{N}\frac{g(\tilde{z})}{A'_{\mu,0}\mathcal{A}'(\tilde{z})}\, .
\end{equation}

\begin{figure}
\center
\includegraphics[width=0.99\textwidth]{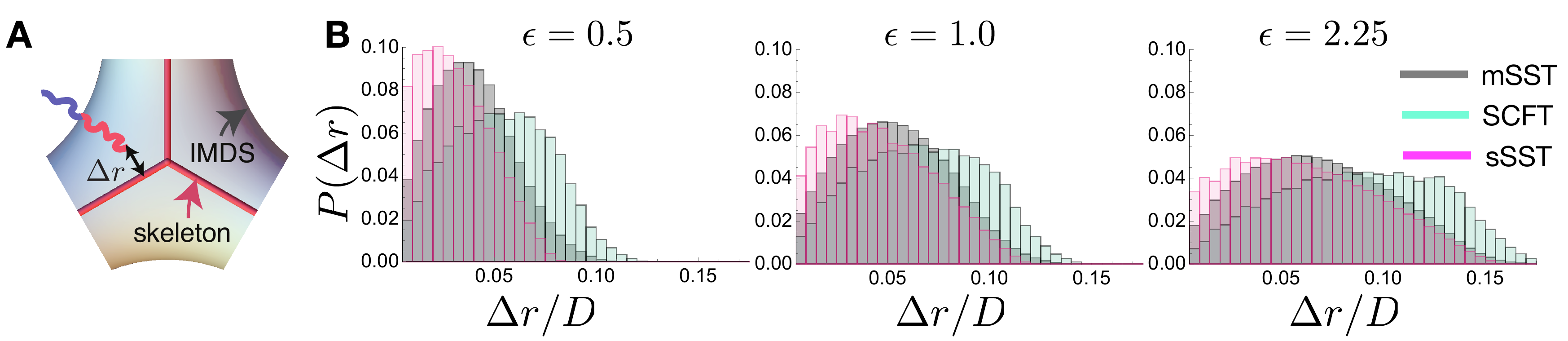}
\caption{\label{fig:end_distributions} (A) Schematics illustrating the distance of end segments spread away from the skeletal graph and in (B) we show distributions of how end segments are spread away from skeletal graph among medial SST, skeletal SST for $f$ along $F(\rm{Lam})=F(\rm{Hex})$ phase boundary and similarly for SCFT.}
\end{figure}

To compare how concentrated A block ends are within the tubular domains, we convert end distributions to the probability density $P(\Delta r)$ of chain ends at $\Delta$ away form the skeletal graph (i.e.~each location a chain end ${\bf x}$ is mapped on to the closest distance to the skeleton).  These distributions are normalized such that $\int_0^\infty {\rm d}(\Delta r) P(\Delta r) =1$.  Figure \ref{fig:end_distributions} compares the distributions for skeletal and medial SST to SCFT predictions ($\chi N = 75$) for three different elastic asymmetries, $\epsilon = 0.5, 1, 2.25$ at the compositions where $F ({\rm Lam}) = F({\rm Hex})$, $f = 0.137, 0.293, 0.496 $ respectively for medial and skeletal SST and for $f = 0.20, 0.32, 0.51$ for SCFT at the same set of $\epsilon$ values.  We note, in general, that free ends are systematically (if only marginally) more concentrated towards the skeletal graph in the skeletal SST relative to the medial SST, consistent with the fact that terminal ends in the former packing are localized along the graph, while in the latter, terminal ends spread out over weblike surface that extends away from the skeleton.  We also, note that ends in $\chi N = 75$ SCFT results are always systematically skewed to large $\Delta r$ than SST predictions, presumably a signature entropic corrections at this finite degree of segregation (e.g.~chain backfolding in the brush).  Altogether all three distributions are shown to shift to large spread from the skeletal graph for increasing $\epsilon$ and $f$ values, consistent with the thickening of the tubular domain with increasing composition.

\section{End-zone corrected results}

\label{sec: eez}

The focus of our presentation relied on the PBT for evaluating the stretching free energy of brush domains, yet it is well known that PBT underestimates the true free energy of convex (outwardly splaying) brushes, as it implicitly allows for a non-positive chain end distribution~\cite{Semenov1985}. The key advantage of the PBT is of course its analytical transparency and geometrically intuitive form eq.~(\ref{eq:PBT}).  Here, we verify that at least for the purposes of modeling the SST thermodynamics of diblocks near to the boundaries between Hex, DG and Lam phase, the PBT is a sufficiently good approximation that corrections due to so-called ``end-zone exclusion" (EEZ) effects can be neglected.

We apply the results of a recent formulation of the exact SST of molten brushes accounting for the effects of an EEZ for brushes of arbitrary shape~\cite{Dimitriyev2021Arxiv}.  This approach builds on a previous self-consistent integral approach by Belyi~\cite{Belyi2004}.
Within an EEZ (close to the anchoring surface at $z=0$), there are no free chain ends; it is a region where the chain end distribution $g(z)$ evaluates to 0.  The presence or absence of an EEZ is controlled by the  effect of curvature coefficients $\tilde{H}$ and $\tilde{K}$ in the area distribution $A(z)=A_0(1+2 \tilde{H} z + \tilde{K} z^2 )$.  Notably, for concave domains for which $\tilde{H}\leq 0$ it can be shown that the EEZ vanishes, and is only non-zero for convex domains, when $\tilde{H}>0$, e.g.~for the outer brushes of spheres and cylinders, {\it and} the outer matrix domain of the DG phase.  Specifically, the magnitude of the EEZ and its corrections to the PBT are controlled by the ratio of these curvatures to the brush thickness, in this case $\tilde{H} h_{\rm B}$ and $\tilde{K} h^2_{\rm B}$.  In ref.~\cite{Dimitriyev2021Arxiv} is was shown that   
the predictions of PBT for $H>0$ are still accurate, with EEZ-corrected free energy increases of order $1\%$ or less appearing for $\tilde{H}h_{\rm B} \lesssim 0.8$; although significant increases on the order of $10\%$ do appear for curvatures $\tilde{H}h_{\rm B} \gtrsim 1.6$.  

To assess the importance of EEZ-corrections for the convex domains on the central conclusions of this study, we also recalculated the free energies Hex and DG using the exact results available from \cite{Dimitriyev2021Arxiv}.  In short, from that work we derived $\Delta F(\tilde{H}h_{\rm B},\tilde{K}h^2_{\rm B}) $ the fractional correction to the PBT stretching free energy due to the EEZ.  Given a local edge $\mu$ with corresponding values of local curvature coefficients $\tilde{H}h_{\rm B}$ and $\tilde{K}h^2_{\rm B}$ we simply multiply the 2nd moments of the region $I_{\mu, {\rm B}}$ by $\big[1 + \Delta F(\tilde{H}h_{\rm B},\tilde{K}h^2_{\rm B}) \big]$, and the remainder of the variational calculation for Hex and DG proceeds as described for the PBT.

\begin{figure}
\center
\includegraphics[width=.9\textwidth]{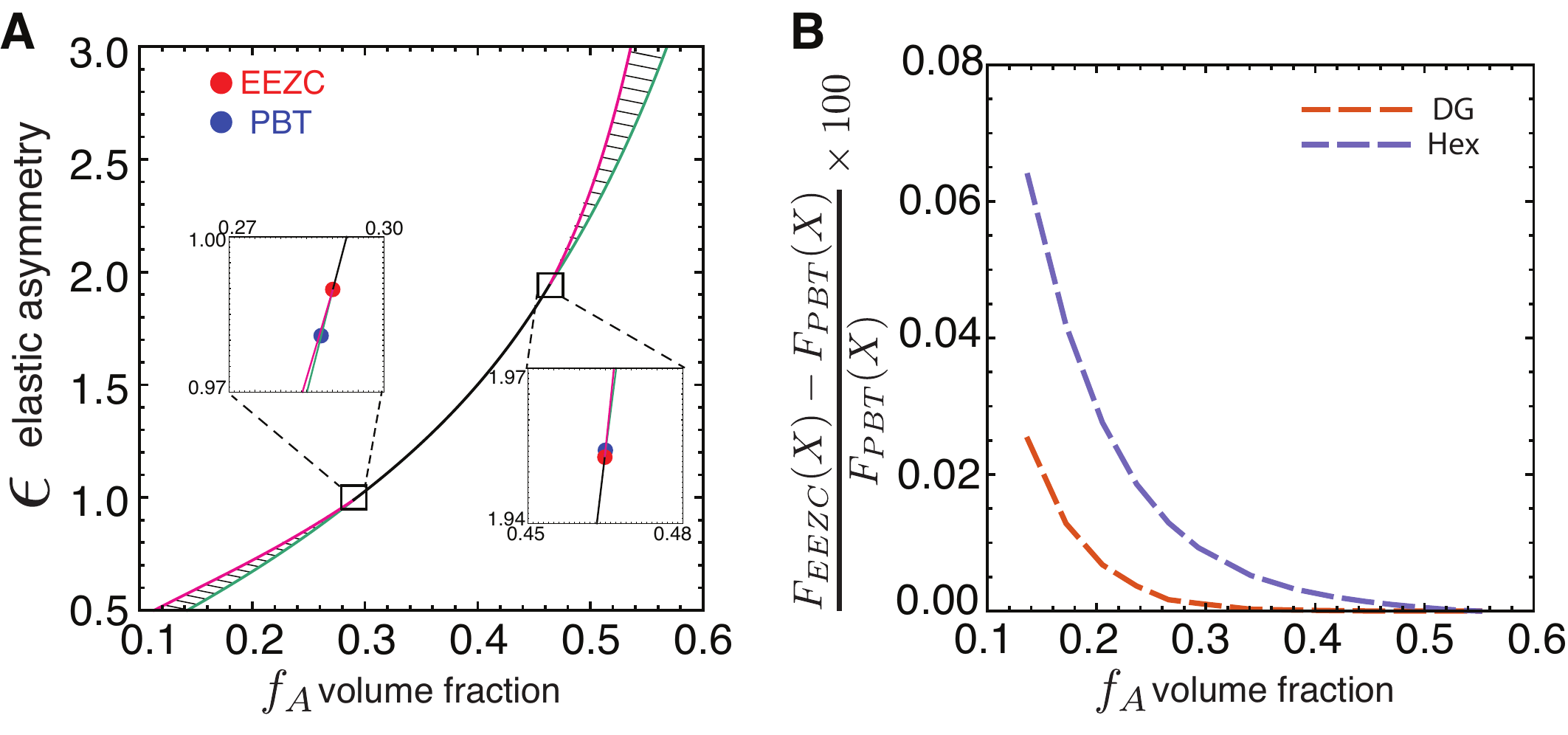}
\caption{\label{fig:eez_correction}(A) Phase boundaries with end-exclusion zone corrections (EEZC) and both insets highlight in triple points with only PBT and with EEZC (B) Relative increase in free energy per chain for DG and Hex are plotted due to EEZC along $F(\rm{Lam})=F(\rm{Hex})$ phase boundary. }
\end{figure}

The change in the phase diagram due to EEZ-corrected stretching free energies is shown in Fig.~\ref{fig:eez_correction}A.  At the scale of the original phase diagram, the boundaries between competing Lam, Hex and DG phases are indistinguishable from the PBT case, hence the overall effect of these corrections (which slightly raise $F({\rm Hex})$ and  $F({\rm DG})$) is minimal for the predicted phase boundaries.  For example, zooming in on the triple points in Fig.~\ref{fig:eez_correction}A, we see the that the EEZ-corrections only displace the lower triple point from the PBT prediction $(f=0.288, \epsilon =0.981)$ to $(f=0.290, \epsilon =0.990)$, corrections less that $10^{-2}$.  

The reason for these very marginal corrections to the phase boundaries is shown in Fig.~\ref{fig:eez_correction}B, where we plot the overall fractional change in the SST free energy for Hex and DG phase along the line where $F({\rm Lam})=F({\rm Hex})$ (based on the PBT approximation).  Because the outer block stretching contributes ultimately only a (small) part of the total free energy of these structures and the EEZ-corrections to the matrix stretching is modest ($\lesssim 1\%$) for the relevant range of IMDS curvatures, the total change in the free energies due to this correction is very small in comparison to the relevant differences in free energy between these phases as a function of $f$.  However, we note that because the IMDS shapes of DG has negative Gaussian curvature, it is effectively ``less convex'' than the Hex phase at the corresponding points in composition space, leading to markedly reduced EEZ-corrections to its free energy.  Hence, we observe a systematic, if modest, shift of these corrections to slightly increase the window of DG stability relative to Hex.

\bibliography{SI_RevTex_MedialSST}